\documentclass[pre,eqsecnum,preprint,showkeys,preprintnumbers,amssymb,amsfonts]{revtex4}

\newcommand{\beq}{\begin{equation}}
\newcommand{\eeq}{\end{equation}}
\newcommand{\beqs}{\begin{eqnarray}}
\newcommand{\eeqs}{\end{eqnarray}}

\begin{document}

\title[Spanning Trees on the 2D Lattices with More Than One Type of Vertex]{Spanning Trees on the Two-Dimensional Lattices with More Than One Type of Vertex} 

\author{Shu-Chiuan Chang$^{a,b}$}
\email{scchang@mail.ncku.edu.tw}

\affiliation{(a) \ Department of Physics \\
National Cheng Kung University \\
Tainan 70101, Taiwan}

\affiliation{(b) \ Physics Division \\
National Center for Theoretical Science \\
National Taiwan University \\
Taipei 10617, Taiwan} 

\begin{abstract}

For a two-dimensional lattice $\Lambda$ with $n$ vertices, the number of spanning trees $N_{ST}(\Lambda)$ grows asymptotically as $\exp(n z_\Lambda)$ in the thermodynamic limit. We present exact integral expression and numerical value for the asymptotic growth constant $z_\Lambda$ for spanning trees on various two-dimensional lattices with more than one type of vertex given in \cite{Okeeffe}. An exact closed-form expression for the asymptotic growth constant is derived for net 14, and the asymptotic growth constants of net 27 and the triangle lattice have the simple relation $z_{27} = (z_{tri}+\ln 4)/4$. Some integral identities are also obtained. 


\end{abstract}

\maketitle

\section{Introduction}

The enumeration of the number of spanning trees $N_{ST}(G)$ on the graph $G$ was first considered by Kirchhoff in the analysis of electric circuits \cite{kirchhoff}. It is
a problem of fundamental interest in mathematics \cite{bbook,welsh,burton93,lyons05} and physics \cite{temperley,wu77}. The number of spanning trees is closely related to the partition function of the $q$-state Potts model in statistical mechanics \cite{fk,wurev}. There are several ways to calculate $N_{ST}(G)$, including as
a determinant of the Laplacian matrix of $G$ and as a special case of the Tutte
polynomial of $G$ \cite{bbook}.  Some recent studies on the enumeration of spanning trees and the calculation of their asymptotic growth constants on regular lattices were carried out in \cite{tzengwu,sw,std,sti}. 
In this paper we shall present exact integrals for the asymptotic growth constant for spanning trees on several two-dimensional lattices given in \cite{Okeeffe}. It was shown in \cite{sti} that integral identities can be obtained with different choice of unit cells, where, in most cases, the sizes of the unit cells are different. Here we shall show that an integral identity can be obtained with the same unit cell but different choice of directions in the calculation. 

\section{Background and Method}

We briefly recall some definitions and background on spanning trees and the
calculation method that we use \cite{bbook,fh}.
Let $G=(V,E)$ denote a connected graph (without loops) with vertex (site) and
edge (bond) sets $V$ and $E$. We will only consider simple graphs without multiple edges.  Let $n=v(G)=|V|$ be the number of vertices and
$e(G)=|E|$ the number of edges in $G$.  A spanning subgraph $G^\prime$ is a
subgraph of $G$ with $v(G^\prime) = |V|$, and a tree is a connected graph
with no circuits. It follows that a spanning tree is a spanning subgraph of $G$ that is a tree and hence $e(G') = n-1$. The degree or coordination number $k_i$ of a vertex
$v_i \in V$ is the number of edges attached to it. A $k$-regular graph is a graph with the property that each of its vertices has the same degree $k$. Two vertices are adjacent if they are connected by an edge in $E$. In general, one can associate an edge weight $x_{ij}$ to each edge connecting adjacent vertices $v_i$ and $v_j$ (see, for example \cite{tzengwu}). For simplicity, all edge weights are set to one throughout this paper.
The adjacency matrix $A(G)$ of $G$ is the $n \times n$ matrix with elements $A(G)_{ij}=1$ if $v_i$ and
$v_j$ are adjacent and zero otherwise. The Laplacian matrix $Q(G)$ is the $n
\times n$ matrix with element $Q(G)_{ij}=k_i\delta_{ij}-A(G)_{ij}$.  One of the
eigenvalues of $Q(G)$ is always zero; let us denote the rest as $\lambda(G)_i$,
$1 \le i \le n-1$.  A basic theorem is that $N_{ST}(G) =
(1/n)\prod_{i=1}^{n-1} \lambda(G)_i$ \cite{bbook}.    
For a $d$-dimensional lattice $\Lambda$ with $d \ge 2$ in
the thermodynamic limit, $N_{ST}(\Lambda)$ grows exponentially with $n$ as 
$n \to \infty$; that is, there exists a constant $z_\Lambda$ such that 
$N_{ST}(\Lambda) \sim \exp(n z_\Lambda)$ as $n \to \infty$.  The constant 
describing this exponential growth is thus given by \cite{burton93,lyons05}
\beq
z_{\Lambda} = \lim_{n \to \infty} \frac{1}{n} \ln \Big [ N_{ST}(\Lambda) \Big ] \ , 
\label{zdef}
\eeq
where $\Lambda$, when used as a subscript in this manner, implicitly refers to
the thermodynamic limit of the lattice $\Lambda$. 

A regular $d$-dimensional lattice is comprised of repeated unit cells, each containing $\nu$ vertices.
Define $a(\tilde n,\tilde n')$ as the $\nu \times \nu$ matrix describing the
adjacency of the vertices of the unit cells $\tilde
n$ and $\tilde n'$, the elements of which are given by $a(\tilde n,\tilde
n')_{ij}=1$ if $v_i \in \tilde n$ is adjacent to $v_j \in \tilde n'$ and zero
otherwise. Although the number of spanning trees $N_{ST}(\Lambda)$ depends on the boundary conditions imposed as shown in \cite{tzengwu}, the asymptotic growth constant $z_{\Lambda}$ is not sensitive to them.
For simplicity, let us consider a given lattice having periodic boundary conditions. Using the resultant translational symmetry for the spanning trees, we have $a(\tilde n, \tilde n')= a(\tilde n- \tilde n')$, and we can therefore write $a(\tilde n)=a(\tilde n_1,\cdots,\tilde n_d)$ for a $d$-dimensional lattice.  Generalizing the method derived in \cite{sw} for lattices which are not $k$-regular, $N_{ST}(\Lambda)$ and $z_\Lambda$ can be calculated in terms of a matrix $M_\Lambda$, which is determined by these $a(\tilde n)$, defined as
\beq
M_\Lambda(\theta_1,\cdots,\theta_d) = M_\Lambda^\prime - \sum_{\tilde n} 
a(\tilde n) e^{i \tilde n \cdot \Theta } \ ,
\label{mmatrix}
\eeq
where $M_\Lambda^\prime$ is the diagonal matrix whose diagonal elements are the degrees $k_i$ of the vertices in the unit cell and $\Theta$ stands for the
$d$-dimensional vector $(\theta_1,\cdots,\theta_d)$.  Then \cite{burton93,sw} 
\beq
z_\Lambda =  {1\over \nu }\int_{-\pi}^\pi 
\biggl [ \prod_{j=1}^d {d\theta_j \over {2\pi}} \biggr ] 
\ln[D_\Lambda(\theta_1,\cdots,\theta_d)] \ , 
\label{zint}
\eeq
where $D_\Lambda(\theta_1,\cdots,\theta_d)=\det (M_\Lambda(\theta_1,\cdots,\theta_d))$ is the determinant of the matrix $M_\Lambda$. Notice that the calculation is not sensitive to the order of the vertex labeling and the choice of the directions $\theta_j$, $1 \le j \le d$. 

It is well known that there are only three uniform tilings of the plane by using one type of regular polygon in which all vertices are equivalent, or three regular tessellations, namely, the square, triangular and honeycomb lattices. If one allows more than one kind of regular polygons and still requests that all vertices are equivalent, there are eight more lattices, or semi-regular tessellations. These are altogether eleven Archimedean lattices which are all $k$-regular \cite{grunbaum}.  If the restriction that all vertices are equivalent is released, an infinite number of tessellations is possible, even with just two types of regular polygons, not to mention if non-regular polygons are allowed. Even though mathematically it is not possible to cover the plane if regular pentagons or heptagons should be presenst, certain arrangements of atoms involving irregular polygons, including pentagons or heptagons, do occur in real-world alloys and inorganic crystals. In \cite{Okeeffe}, various common tessellations (including the eleven Archimedean lattices), denoted as nets, and their occurrences were given. 

For a lattice $\Lambda$ which is not $k$-regular, it is convenient to introduce an effective coordination number $\kappa_\Lambda$, defined as the average number of edges per vertex, 
\beq
\kappa_\Lambda = \lim_{n(\Lambda) \to \infty} \frac{2e(\Lambda)}{n(\Lambda)} \ .
\eeq
For a $k$-regular lattice, $\kappa = k$.
Furthermore, we know that the number of spanning trees is the same for a planar graph $G$ and its dual $G^*$, and the number of the vertices of $G^*$ is given by the Euler relation $v(G^*)=e(G)-n+1$. It follows that the asymptotic growth constants of $G$ and $G^*$ satisfy the relation \cite{sw, sti}
\beq
z_{G^*} = \frac{z_G}{\kappa/2-1} \ .
\label{zdual}
\eeq 

For a $k$-regular graph $G_k$, a general upper bound for the asymptotic growth constant is $z_{G_k} \le \ln k$ \cite{grimmett}.  A stronger upper bound for $G_k$ with
$k \ge 3$ was derived in \cite{mckay,chungyau} that
\beq
N_{ST}(G_k) \le \Biggl ( \frac{2\ln n}{n k \ln k} \Bigg) (b_k)^n \ ,
\label{nmckay}
\eeq
where
\beq
b_k = \frac{(k-1)^{k-1}}{[k(k-2)]^{\frac{k}{2}-1}} \ . 
\label{ck}
\eeq
By Eq. (\ref{zdef}), this then yields \cite{sw} 
\beq
z_{G_k} \leq \ln(b_k) \ . 
\label{mcybound}
\eeq

\section{Asymptotic Growth Constants}

The asymptotic growth constants $z_\Lambda$ for the eleven Archimedean lattices have been considered by several authors \cite{temperley,wu77,sw,std,sti}. While the relation $z_{hc}=z_{tri}/2$ for the honeycomb and triangular lattices is easy to understand due to the duality \cite{sw} (cf. Eq. (\ref{zdual})), it is non-trivial to have the relations $z_{kag}=(z_{tri}+\ln6)/3$ for the Kagom\'e (equivalently $(3.6.3.6)$) lattice and $z_{(3.12.12)}=(z_{tri}+\ln(15))/6$ for the $(3.12.12)$ lattice given in \cite{sw}. Our main purpose is to calculate the asymptotic growth constants for other common two-dimensional lattices where more than one type of vertex occur. Following the denotation given in \cite{Okeeffe}, we shall quote them as net 12 to net 27, where net 12 to net 17 are tessellations with two or three regular polygons (including triangle, square or hexagon), while net 18 to net 27 are tessellations containing pentagons, heptagons or enneagons. In addition, \cite{Okeeffe} mentions the B net of YCrB$_4$, the B net of Y$_2$LnB$_6$ and a net with $5^2.8$ and $5.8^2$ vertices. Let us denote them as nets 28, 29 and 30, respectively. The figures of these nets are referred to those in \cite{Okeeffe}, and the unit cells chosen for the calculation are shown in Figs. \ref{netfig1}-\ref{netfig3}. Notice that the polygons are regular or not does not affect the number of spanning trees, so the unit cells can be deformed from those in \cite{Okeeffe}. In addition to the numerical values of the asymptotic growth constants, we shall derive an exact closed-form expression for $z_{14}$, and the relation $z_{27} = (z_{tri}+\ln 4)/4$. A few integral identities will be given by choosing different unit cells or directions. We will use the shorthand notations $\alpha = e^{i\theta_1}$ and $\beta = e^{i\theta_2}$ for the elements of the matrix $M_\Lambda$ when the matrix is large. 

\subsection{Nets with regular polygons}

In this subsection ,we consider important nets in which only regular polygons (including triangle, square or hexagon) occur with more than one type of vertex.

\subsubsection{Net 12}

\begin{figure}[htbp]
\unitlength 0.8mm \hspace*{5mm}
\begin{picture}(170,50)
\put(5,0){\line(1,0){10}}
\multiput(5,0)(10,0){2}{\line(-1,2){5}}
\multiput(5,0)(10,0){2}{\line(1,2){5}}
\multiput(0,10)(0,10){2}{\line(1,0){20}}
\multiput(0,10)(10,0){3}{\line(0,1){10}}
\multiput(0,20)(10,0){2}{\line(1,2){5}}
\multiput(10,20)(10,0){2}{\line(-1,2){5}}
\multiput(5,30)(0,10){2}{\line(1,0){10}}
\multiput(5,30)(10,0){2}{\line(0,1){10}}
\multiput(5,0)(10,0){2}{\circle*{2}}
\multiput(0,10)(10,0){3}{\circle*{2}}
\multiput(0,20)(10,0){3}{\circle*{2}}
\multiput(5,30)(10,0){2}{\circle*{2}}
\multiput(5,40)(10,0){2}{\circle*{2}}
\put(17,32){\makebox(0,0){\footnotesize 1}}
\put(17,42){\makebox(0,0){\footnotesize 2}}
\put(3,42){\makebox(0,0){\footnotesize 3}}
\put(3,32){\makebox(0,0){\footnotesize 4}}
\put(-2,22){\makebox(0,0){\footnotesize 5}}
\put(-2,12){\makebox(0,0){\footnotesize 6}}
\put(3,-2){\makebox(0,0){\footnotesize 7}}
\put(17,-2){\makebox(0,0){\footnotesize 8}}
\put(22,12){\makebox(0,0){\footnotesize 9}}
\put(23,22){\makebox(0,0){\footnotesize 10}}
\put(13,22){\makebox(0,0){\footnotesize 11}}
\put(13,12){\makebox(0,0){\footnotesize 12}}
\put(10,-7){\makebox(0,0){$(a)$}}

\multiput(40,0)(0,10){6}{\line(1,0){60}}
\multiput(40,0)(10,0){7}{\line(0,1){50}}
\multiput(40,10)(10,0){2}{\line(1,1){10}}
\multiput(40,30)(10,0){2}{\line(1,1){10}}
\multiput(60,10)(0,20){3}{\line(1,-1){10}}
\multiput(70,10)(0,20){3}{\line(1,-1){10}}
\multiput(80,10)(10,0){2}{\line(1,1){10}}
\multiput(80,30)(10,0){2}{\line(1,1){10}}
\multiput(60,20)(10,0){4}{\circle*{2}}
\multiput(60,30)(10,0){4}{\circle*{2}}
\put(88,32){\makebox(0,0){\footnotesize 1}}
\put(78,32){\makebox(0,0){\footnotesize 2}}
\put(68,32){\makebox(0,0){\footnotesize 3}}
\put(58,32){\makebox(0,0){\footnotesize 4}}
\put(62,18){\makebox(0,0){\footnotesize 5}}
\put(72,18){\makebox(0,0){\footnotesize 6}}
\put(82,18){\makebox(0,0){\footnotesize 7}}
\put(92,18){\makebox(0,0){\footnotesize 8}}
\put(70,-7){\makebox(0,0){$(b)$}}

\multiput(120,0)(0,20){3}{\line(1,0){50}}
\multiput(120,0)(20,0){3}{\line(1,2){10}}
\multiput(130,0)(20,0){3}{\line(-1,2){10}}
\multiput(120,20)(20,0){3}{\line(1,2){10}}
\multiput(130,20)(20,0){3}{\line(-1,2){10}}
\multiput(120,40)(20,0){3}{\line(1,2){5}}
\multiput(130,40)(20,0){3}{\line(-1,2){5}}
\multiput(140,20)(10,0){2}{\circle*{2}}
\put(145,30){\circle*{2}}
\put(147,30){\makebox(0,0){\footnotesize 1}}
\put(138,22){\makebox(0,0){\footnotesize 2}}
\put(152,22){\makebox(0,0){\footnotesize 3}}
\put(155,-7){\makebox(0,0){$(c)$}}
\end{picture}

\vspace*{15mm}

\begin{picture}(155,60)
\multiput(39,0)(-21,6){2}{\line(1,0){18}}
\multiput(39,12)(-21,6){2}{\line(1,0){24}}
\multiput(45,24)(-21,6){3}{\line(1,0){18}}
\multiput(24,42)(-21,6){2}{\line(1,0){24}}
\multiput(30,54)(-21,6){2}{\line(1,0){18}}
\put(45,36){\line(1,0){12}}
\put(9,24){\line(1,0){12}}
\multiput(57,0)(-21,6){3}{\line(1,2){9}}
\multiput(45,0)(-21,6){2}{\line(1,2){12}}
\multiput(42,30)(-21,6){3}{\line(1,2){9}}
\multiput(30,30)(-21,6){2}{\line(1,2){12}}
\put(51,24){\line(1,2){6}}
\put(9,24){\line(1,2){6}}
\multiput(39,0)(-21,6){2}{\line(-1,2){3}}
\multiput(51,0)(-21,6){2}{\line(-1,2){9}}
\multiput(60,6)(-21,6){3}{\line(-1,2){9}}
\multiput(66,18)(-21,6){4}{\line(-1,2){3}}
\multiput(36,30)(-21,6){2}{\line(-1,2){9}}
\multiput(45,36)(-21,6){2}{\line(-1,2){9}}
\multiput(51,48)(-21,6){2}{\line(-1,2){3}}
\put(57,24){\line(-1,2){9}}
\multiput(30,6)(-3,6){4}{\circle*{2}}
\multiput(36,6)(-6,12){3}{\circle*{2}}
\multiput(39,12)(-3,6){4}{\circle*{2}}
\put(41,10){\makebox(0,0){\footnotesize 1}}
\put(38,20){\makebox(0,0){\footnotesize 2}}
\put(35,24){\makebox(0,0){\footnotesize 3}}
\put(28,32){\makebox(0,0){\footnotesize 4}}
\put(22,30){\makebox(0,0){\footnotesize 5}}
\put(19,26){\makebox(0,0){\footnotesize 6}}
\put(22,16){\makebox(0,0){\footnotesize 7}}
\put(25,12){\makebox(0,0){\footnotesize 8}}
\put(28,4){\makebox(0,0){\footnotesize 9}}
\put(34,4){\makebox(0,0){\footnotesize 10}}
\put(32,16){\makebox(0,0){\footnotesize 11}}
\put(33,-7){\makebox(0,0){$(d)$}}

\multiput(100,0)(5,20){2}{\line(2,1){10}}
\multiput(100,0)(-5,20){2}{\line(-2,1){10}}
\put(100,0){\line(0,1){10}}
\multiput(100,10)(10,-5){2}{\line(1,2){5}}
\multiput(100,10)(-10,-5){2}{\line(-1,2){5}}
\multiput(100,10)(5,10){2}{\line(2,-1){10}}
\multiput(100,10)(-5,10){2}{\line(-2,-1){10}}
\multiput(85,15)(30,0){2}{\line(0,1){10}}
\multiput(95,20)(0,10){2}{\line(1,0){10}}
\multiput(95,20)(10,0){2}{\line(0,1){10}}
\put(85,25){\line(2,1){10}}
\put(115,25){\line(-2,1){10}}
\multiput(100,0)(0,10){2}{\circle*{2}}
\multiput(90,5)(20,0){2}{\circle*{2}}
\multiput(85,15)(0,10){2}{\circle*{2}}
\multiput(115,15)(0,10){2}{\circle*{2}}
\multiput(95,20)(0,10){2}{\circle*{2}}
\multiput(105,20)(0,10){2}{\circle*{2}}
\put(102,-2){\makebox(0,0){\footnotesize 1}}
\put(112,3){\makebox(0,0){\footnotesize 2}}
\put(117,13){\makebox(0,0){\footnotesize 3}}
\put(117,27){\makebox(0,0){\footnotesize 4}}
\put(107,32){\makebox(0,0){\footnotesize 5}}
\put(93,32){\makebox(0,0){\footnotesize 6}}
\put(83,27){\makebox(0,0){\footnotesize 7}}
\put(83,13){\makebox(0,0){\footnotesize 8}}
\put(88,3){\makebox(0,0){\footnotesize 9}}
\put(103,11){\makebox(0,0){\footnotesize 10}}
\put(107,17){\makebox(0,0){\footnotesize 11}}
\put(93,17){\makebox(0,0){\footnotesize 12}}
\put(100,-7){\makebox(0,0){$(e)$}}

\multiput(135,20)(10,0){3}{\line(0,1){10}}
\multiput(145,0)(-10,30){2}{\line(1,2){10}}
\multiput(145,0)(10,30){2}{\line(-1,2){10}}
\multiput(140,10)(0,30){2}{\line(1,0){10}}
\multiput(135,20)(0,10){2}{\line(1,0){20}}
\multiput(140,10)(5,20){2}{\line(1,2){5}}
\multiput(150,10)(-5,20){2}{\line(-1,2){5}}
\multiput(145,0)(0,50){2}{\circle*{2}}
\multiput(140,10)(10,0){2}{\circle*{2}}
\multiput(135,20)(10,0){3}{\circle*{2}}
\multiput(135,30)(10,0){3}{\circle*{2}}
\multiput(140,40)(10,0){2}{\circle*{2}}
\put(143,52){\makebox(0,0){\footnotesize 1}}
\put(138,42){\makebox(0,0){\footnotesize 2}}
\put(133,32){\makebox(0,0){\footnotesize 3}}
\put(133,18){\makebox(0,0){\footnotesize 4}}
\put(138,8){\makebox(0,0){\footnotesize 5}}
\put(143,-2){\makebox(0,0){\footnotesize 6}}
\put(152,8){\makebox(0,0){\footnotesize 7}}
\put(157,18){\makebox(0,0){\footnotesize 8}}
\put(157,32){\makebox(0,0){\footnotesize 9}}
\put(153,42){\makebox(0,0){\footnotesize 10}}
\put(148,28){\makebox(0,0){\footnotesize 11}}
\put(148,22){\makebox(0,0){\footnotesize 12}}
\put(145,-7){\makebox(0,0){$(f)$}}
\end{picture}

\vspace*{15mm}

\begin{picture}(50,70)
\multiput(20,0)(0,70){2}{\line(1,0){10}}
\multiput(20,20)(0,10){2}{\line(1,0){10}}
\multiput(20,0)(15,10){2}{\line(-1,2){5}}
\multiput(30,0)(-15,10){2}{\line(1,2){5}}
\multiput(20,20)(10,0){2}{\line(0,1){10}}
\multiput(20,20)(20,5){2}{\line(-2,1){10}}
\multiput(20,30)(20,-5){2}{\line(-2,-1){10}}
\multiput(10,25)(30,0){2}{\line(0,1){10}}
\multiput(20,30)(-10,5){2}{\line(1,2){5}}
\multiput(20,30)(5,10){2}{\line(-2,1){10}}
\multiput(30,30)(-5,10){2}{\line(2,1){10}}
\multiput(30,30)(10,5){2}{\line(-1,2){5}}
\multiput(0,35)(40,0){2}{\line(1,0){10}}
\multiput(0,35)(40,0){2}{\line(1,2){5}}
\multiput(10,35)(40,0){2}{\line(-1,2){5}}
\multiput(5,45)(0,10){2}{\line(1,0){10}}
\multiput(5,45)(10,0){2}{\line(0,1){10}}
\multiput(35,45)(0,10){2}{\line(1,0){10}}
\multiput(35,45)(10,0){2}{\line(0,1){10}}
\put(5,55){\line(1,2){5}}
\put(45,55){\line(-1,2){5}}
\multiput(15,55)(10,5){2}{\line(-1,2){5}}
\multiput(15,55)(-5,10){2}{\line(2,1){10}}
\multiput(35,55)(-10,5){2}{\line(1,2){5}}
\multiput(35,55)(5,10){2}{\line(-2,1){10}}
\multiput(20,0)(10,0){2}{\circle*{2}}
\multiput(15,10)(20,0){2}{\circle*{2}}
\multiput(20,20)(10,0){2}{\circle*{2}}
\multiput(20,30)(10,0){2}{\circle*{2}}
\multiput(10,25)(30,0){2}{\circle*{2}}
\multiput(0,35)(10,0){2}{\circle*{2}}
\multiput(40,35)(10,0){2}{\circle*{2}}
\multiput(25,40)(0,20){2}{\circle*{2}}
\multiput(5,45)(10,0){2}{\circle*{2}}
\multiput(35,45)(10,0){2}{\circle*{2}}
\multiput(5,55)(10,0){2}{\circle*{2}}
\multiput(35,55)(10,0){2}{\circle*{2}}
\multiput(10,65)(30,0){2}{\circle*{2}}
\multiput(20,70)(10,0){2}{\circle*{2}}
\put(52,37){\makebox(0,0){\footnotesize 1}}
\put(47,47){\makebox(0,0){\footnotesize 2}}
\put(47,57){\makebox(0,0){\footnotesize 3}}
\put(42,67){\makebox(0,0){\footnotesize 4}}
\put(32,72){\makebox(0,0){\footnotesize 5}}
\put(18,72){\makebox(0,0){\footnotesize 6}}
\put(8,67){\makebox(0,0){\footnotesize 7}}
\put(3,57){\makebox(0,0){\footnotesize 8}}
\put(3,47){\makebox(0,0){\footnotesize 9}}
\put(-3,37){\makebox(0,0){\footnotesize 10}}
\put(7,33){\makebox(0,0){\footnotesize 11}}
\put(7,23){\makebox(0,0){\footnotesize 12}}
\put(16,19){\makebox(0,0){\footnotesize 13}}
\put(12,8){\makebox(0,0){\footnotesize 14}}
\put(17,-2){\makebox(0,0){\footnotesize 15}}
\put(33,-2){\makebox(0,0){\footnotesize 16}}
\put(38,8){\makebox(0,0){\footnotesize 17}}
\put(34,19){\makebox(0,0){\footnotesize 18}}
\put(43,23){\makebox(0,0){\footnotesize 19}}
\put(43,33){\makebox(0,0){\footnotesize 20}}
\put(32,46){\makebox(0,0){\footnotesize 21}}
\put(32,54){\makebox(0,0){\footnotesize 22}}
\put(25,57){\makebox(0,0){\footnotesize 23}}
\put(18,54){\makebox(0,0){\footnotesize 24}}
\put(18,46){\makebox(0,0){\footnotesize 25}}
\put(25,43){\makebox(0,0){\footnotesize 26}}
\put(19,33){\makebox(0,0){\footnotesize 27}}
\put(31,33){\makebox(0,0){\footnotesize 28}}
\put(25,-7){\makebox(0,0){$(g)$}}
\end{picture}

\vspace*{5mm}
\caption{\footnotesize{(a) A unit cell of net 12. (b) Net 13. (c) Net 14. (d) Net 15. (e) A unit cell of net 16(a). (f) A unit cell of net 16(b). (g) A unit cell of net 17. Vertices within a unit cell are labeled.}} 
\label{netfig1}
\end{figure}
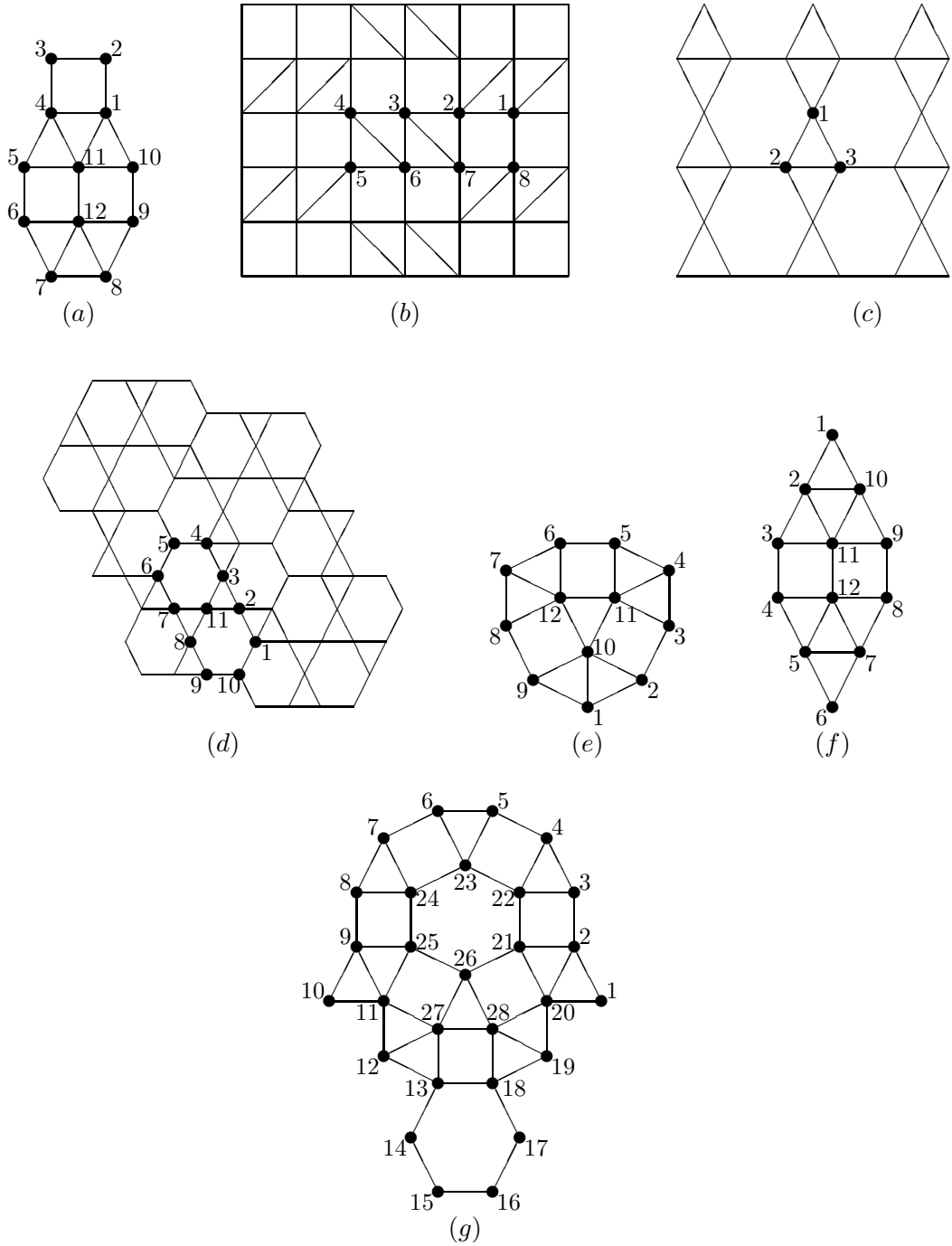

Net 12 is the combinations of $3^2.4.3.4$ and $3^3.4^2$ vertices. A primitive unit cell contains twelve vertices $\nu_{12}=12$, and the coordination number is $k_{12}=5$. By the choice of the unit cell and vertex labeling shown in Fig. \ref{netfig1} (a), we have
\beqs
M_{12}(\theta_1,\theta_2)
& = & \left( \begin{array}{cccccccccccc}
5 & -1 & 0 & -1 & 0 & -\alpha & 0 & 0 & 0 & -1 & -1 & 0 \\
-1 & 5 & -1 & 0 & -\alpha & -\alpha & 0 & -\alpha\beta & 0 & 0 & 0 & 0 \\
0 & -1 & 5 & -1 & 0 & 0 & -\alpha\beta & 0 & -\beta & -\beta & 0 & 0 \\
-1 & 0 & -1 & 5 & -1 & 0 & 0 & 0 & -\beta & 0 & -1 & 0 \\
0 & -\frac{1}{\alpha} & 0 & -1 & 5 & -1 & 0 & -\beta & 0 & 0 & -1 & 0 \\ -\frac{1}{\alpha} & -\frac{1}{\alpha} & 0 & 0 & -1 & 5 & -1 & 0 & 0 & 0 & 0 & -1 \\
0 & 0 & -\frac{1}{\alpha\beta} & 0 & 0 & -1 & 5 & -1 & 0 & -\frac{1}{\alpha} & 0 & -1 \\
0 & -\frac{1}{\alpha\beta} & 0 & 0 & -\frac{1}{\beta} & 0 & -1 & 5 & -1 & 0 & 0 & -1 \\
0 & 0 & -\frac{1}{\beta} & -\frac{1}{\beta} & 0 & 0 & 0 & -1 & 5 & -1 & 0 & -1 \\
-1 & 0 & -\frac{1}{\beta} & 0 & 0 & 0 & -\alpha & 0 & -1 & 5 & -1 & 0 \\
-1 & 0 & 0 & -1 & -1 & 0 & 0 & 0 & 0 & -1 & 5 & -1 \\
0 & 0 & 0 & 0 & 0 & -1 & -1 & -1 & -1 & 0 & -1 & 5 
\end{array} \right ) \ . 
\eeqs
The determinant can be calculated to be
\beqs
D_{12}(\theta_1,\theta_2)
& = & 16 \{1650732 - 680016 (\cos \theta_1 + \cos \theta_2) + 10151 (\cos^2 \theta_1 + \cos^2 \theta_2) \cr\cr
& & - 300022 \cos \theta_1 \cos \theta_2 - (\cos^3 \theta_1 + \cos^3 \theta_2) - 5567 \cos \theta_1 \cos \theta_2 (\cos \theta_1 + \cos \theta_2) \cr\cr & & + 158 \cos^2 \theta_1 \cos^2 \theta_2 - \cos \theta_1 \cos \theta_2 (\cos^2 \theta_1 + \cos^2 \theta_2) \} \ ,
\eeqs
such that the numerical evaluation gives
\beq
z_{12} = \frac{1}{12} \int_{-\pi}^{\pi} \frac{d\theta_1}{2\pi} \int_{-\pi}^{\pi} \frac{d\theta_2}{2\pi}
\ln \Big [ D_{12}(\theta_1,\theta_2) \Big ] = 1.409737903756929...
\eeq

\subsubsection{Net 13}

Net 13 is obtained by intergrowth of $3^2.4.3.4$ and $3^3.4^2$ vertices. It can be constructed by starting with the square lattice and adding appropriate diagonal edges as shown in Fig. \ref{netfig1} (b), such that the coordination number is $k_{13}=5$. Taking three contiguous squares where two of them contains diagonal edges as a unit cell with $\nu_{13}=8$, we have
\beq
M_{13}(\theta_1,\theta_2)
= \left( \begin{array}{cccccccc}
5 & -1 & 0 & -\alpha & -\alpha\beta & 0 & 0 & -1-\beta \\ 
-1 & 5 & -1 & 0 & 0 & 0 & -1-\beta & -\beta \\
0 & -1 & 5 & -1 & 0 & -1-\beta & -1 & 0 \\
-\frac{1}{\alpha} & 0 & -1 & 5 & -1-\beta & -1 & 0 & 0 \\
-\frac{1}{\alpha\beta} & 0 & 0 & -1-\frac{1}{\beta} & 5 & -1 & 0 & -\frac{1}{\alpha} \\
0 & 0 & -1-\frac{1}{\beta} & -1 & -1 & 5 & -1 & 0 \\
0 & -1-\frac{1}{\beta} & -1 & 0 & 0 & -1 & 5 & -1 \\
-1-\frac{1}{\beta} & -\frac{1}{\beta} & 0 & 0 & -\alpha & 0 & -1 & 5
\end{array} \right ) \ . 
\eeq
The determinant can be calculated to be
\beqs
D_{13}(\theta_1,\theta_2)
& = & 4\{25560 - 2328\cos \theta_1 - 25872 \cos \theta_2 + \cos^2 \theta_1 + 5397 \cos^2 \theta_2 \cr\cr
& & - 2158 \cos \theta_1 \cos \theta_2 - 288 \cos^3 \theta_2 - 312 \cos \theta_1 \cos^2 \theta_2 + 4 \cos^4 \theta_2 \cr\cr
& & - 4 \cos \theta_1 \cos^3 \theta_2 \} \ ,
\eeqs
such that
\beq
z_{13} = \frac{1}{8} \int_{-\pi}^{\pi} \frac{d\theta_1}{2\pi} \int_{-\pi}^{\pi} \frac{d\theta_2}{2\pi}
\ln \Big [ D_{13}(\theta_1,\theta_2) \Big ] = 1.409133286424679... 
\eeq
The $z$ values of the variants of net 13 shown in Fig. 15 of \cite{Okeeffe} are close to this, and are not given here to save space.

\subsubsection{Net 14}

Net 14 is a simple combination of hexagons and triangles as shown in Fig. \ref{netfig1} (c), such that the coordination number is $k_{14}=4$. Taking a triangle as a unit cell with $\nu_{14}=3$, we have
\beq
M_{14}(\theta_1,\theta_2) = \left( \begin{array}{ccc}
4 & -1-e^{i\theta_2} & -1-e^{i\theta_2} \\
-1-e^{-i\theta_2} & 4 & -1-e^{-i\theta_1} \\
-1-e^{-i\theta_2} & -1-e^{i\theta_1} & 4
\end{array} \right ) \ .
\eeq
The determinant can be calculated to be
\beq
D_{14}(\theta_1,\theta_2) = 4 (9 - 3\cos \theta_1 - 5\cos \theta_2 - \cos \theta_1 \cos \theta_2) \ ,
\eeq
such that
\beqs
z_{14} & = & \frac{1}{3} \int_{-\pi}^{\pi} \frac{d\theta_1}{2\pi} \int_{-\pi}^{\pi} \frac{d\theta_2}{2\pi}
\ln \Big [ D_{14}(\theta_1,\theta_2) \Big ] \cr\cr
& = & \frac{\ln 4}{3} + \frac{1}{3} \int_0^{\pi} \frac{d\theta_1}{\pi} \ln \Bigl [ \frac{9-3\cos \theta_1 + \sqrt{8(1-\cos \theta_1)(7-\cos \theta_1)}}{2} \Bigr ] \cr\cr
& = & \frac{\ln 2}{3} + \frac{2}{3} \int_0^{\pi} \frac{d\theta_1}{\pi} \ln
\Big ( 2\sin \frac{\theta_1}{2} + \sqrt{6+2\sin^2 \frac{\theta_1}{2}} \Big ) \ .
\eeqs
An exact closed-form expression for this integral can be derived as follows.  After changing the variable $\theta_1 = 2\theta$, we have
\beqs
z_{14} & = & \ln 2 + \frac{4}{3\pi} \int_0^{\pi /2} d\theta \ln \Big ( \sin \theta + \sqrt{\frac{3}{2}+\frac{1}{2}\sin^2 \theta} \Big ) \cr\cr
& = & \ln 2 + \frac{4}{3} I \Bigl ( \frac{3}{2}, \frac{1}{2} \Bigr ) \ ,
\label{z14}
\eeqs
where
\beq
I(a,b) = \frac{1}{\pi} \int_0^{\pi/2} d\theta \ \ln \Bigl ( \sin \theta 
+ \sqrt{a+b \sin^2\theta } \ \Bigr ) 
\label{iint}
\eeq
and we consider $0 \le b < 1 \le a$. For $b=0$, it can be shown that
\beqs
I(a,0) & = & \frac{1}{\pi} \int_{0}^{\pi/2} d\theta \ \ln (\sin \theta +\sqrt{a}) \cr\cr
& = & -\frac{1}{2} \ln [2(\sqrt{a}+\sqrt{a-1})] + \frac{2}{\pi} {\rm Ti}_2 (\sqrt{a}+\sqrt{a-1}) \ ,
\label{izero}
\eeqs
where Ti$_2(x)$ is the inverse tangent integral \cite{invtan}, 
\beqs
{\rm Ti}_2(x) & = & \int_0^x \frac{\tan^{-1} t}{t} \ dt 
= \sum_{k=1}^\infty (-1)^{k-1} \frac{x^{2k-1}}{(2k-1)^2} \cr\cr
& = & \frac{1}{2i} [{\rm Li}_2(ix) - {\rm Li}_2(-ix)] \ .
\eeqs
Here dilogarithm ${\rm Li}_2(z)$ is defined by
\beq
{\rm Li}_2(z) = \sum_{k=1}^\infty \frac{z^k}{k^2} = \int_z^0 \frac{\ln(1-t)}{t} dt \ .
\eeq
Taking the derivative with respect to $b$ then performing the integral over $\theta$ in Eq. (\ref{iint}), we get
\beq
\frac{d}{db}I(a,b) = \frac{1}{2\pi (1-b)} \Bigl ( \frac{1}{\sqrt{b}} \tan^{-1} \sqrt{\frac{b}{a}} + \sqrt{\frac{a}{a+b-1}} \tan^{-1} \sqrt{a+b-1} \Bigr ) - \frac{1}{4(1-b)} \ .
\label{iprime}
\eeq
It can be integrated to give
\beqs
\lefteqn{I(a,b) = I(a,0) + \int_{0}^{b} I'(a,x) dx} \cr\cr
& = & \frac{1}{2} \ln \Bigl [ \frac{\sqrt{1-b}}{2(\sqrt{a}+\sqrt{a-1})} \Bigr ] + \frac{2}{\pi} {\rm Ti}_2 (\sqrt{a}+\sqrt{a-1}) + \frac{1}{\pi} \Bigl ( \tanh^{-1} \sqrt{b} \Bigr ) \Bigl ( \tan^{-1} \frac{1}{\sqrt{a}} \Bigr ) \cr\cr
& & + \frac{1}{\pi} \Bigl ( \tanh^{-1} \sqrt{1-\frac{1-b}{a}} - \tanh^{-1} \sqrt{1-\frac{1}{a}} \Bigr ) \tan^{-1} \sqrt{a} \cr\cr
& & + \frac{1}{2\pi} \sum_{k=1}^\infty \Bigl [ \bigl ( \frac{\sqrt{a-1}-\sqrt{a}}{\sqrt{a-1}+\sqrt{a}} \bigr )^k - \bigl ( \frac{\sqrt{a+b-1}-\sqrt{a}}{\sqrt{a+b-1}+\sqrt{a}} \bigr )^k + \bigl ( \frac{1-\sqrt{b}}{1+\sqrt{b}} \bigr )^k -1 \Bigr ] \frac{\sin(k\phi)}{k^2} \cr\cr
& = & \frac{1}{4} \ln \Bigl [ \frac{(1-b)(1+\sqrt{b})}{4(2a-1+2\sqrt{a(a-1)})(1-\sqrt{b})} \Bigr ] + \frac{2}{\pi} {\rm Ti}_2 (\sqrt{a}+\sqrt{a-1}) \cr\cr
& & + \frac{1}{\pi} \Bigl ( \tanh^{-1} \sqrt{1-\frac{1-b}{a}} - \tanh^{-1} \sqrt{1-\frac{1}{a}} - \tanh^{-1} \sqrt{b} \Bigr ) \tan^{-1} \sqrt{a} 
\cr\cr
& & + \frac{1}{4\pi i} \Bigl [ {\rm Li}_2 \bigl ( \frac{\sqrt{a-1}-\sqrt{a}}{\sqrt{a-1}+\sqrt{a}} e^{i\phi} \bigr ) - {\rm Li}_2 \bigl ( \frac{\sqrt{a-1}-\sqrt{a}}{\sqrt{a-1}+\sqrt{a}} e^{-i\phi} \bigr ) \cr\cr
& & - {\rm Li}_2 \bigl ( \frac{\sqrt{a+b-1}-\sqrt{a}}{\sqrt{a+b-1}+\sqrt{a}} e^{i\phi} \bigr ) + {\rm Li}_2 \bigl ( \frac{\sqrt{a+b-1}-\sqrt{a}}{\sqrt{a+b-1}+\sqrt{a}} e^{-i\phi} \bigr ) \cr\cr
& & + {\rm Li}_2 \bigl ( \frac{1-\sqrt{b}}{1+\sqrt{b}} e^{i\phi} \bigr ) - {\rm Li}_2 \bigl ( \frac{1-\sqrt{b}}{1+\sqrt{b}} e^{-i\phi} \bigr ) - {\rm Li}_2 \bigl ( e^{i\phi} \bigr ) + {\rm Li}_2 \bigl ( e^{-i\phi} \bigr ) \Bigr ] \ ,
\label{iab}
\eeqs
where $\phi = \tan^{-1} [2\sqrt{a}/(1-a)]$ and $\pi/2 \le \phi < \pi$. When $b$ is set to zero, it is clear that Eq. (\ref{iab}) reduces to Eq. (\ref{izero}). We notice that when $a$ is set to one, the expression for $I(a=1,b)$ can be simplified, using the identity that Ti$_2(1)$ is equal to the Catalan constant $C = \sum_{n=0}^\infty (-1)^n (2n+1)^{-2}=0.9159655941772190...$, as
\beq
I(1,b) = \frac{1}{4} \ln \Bigl ( \frac{1-b}{4} \Bigr ) + \frac{C}{\pi} + \frac{1}{2} \tanh^{-1} \sqrt{b} + \frac{1}{\pi} {\rm Ti}_2 \Bigl ( \frac{1-\sqrt{b}}{1+\sqrt{b}} \Bigr ) \ ,
\eeq
which is equivalent to Eq. (28) of \cite{std}.
Evaluating $I(a,b)$ in Eq. (\ref{iab}) at $a=3/2$, $b=1/2$ and substituting into Eq. (\ref{z14}), we obtain the exact closed-form expression
\beqs
z_{14} & = & \frac{1}{3} \ln \Bigl ( \frac{3+2\sqrt{2}}{2+\sqrt{3}} \Bigr ) + \frac{8}{3\pi} {\rm Ti}_2 \Bigl ( \frac{\sqrt{3}+1}{\sqrt{2}} \Bigr ) + \frac{4}{3\pi} \Bigl ( \tan^{-1} \sqrt{\frac{3}{2}} \Bigr ) \Bigl ( \tanh^{-1} \frac{4\sqrt{3}-\sqrt{6}-7}{\sqrt{3}-2\sqrt{6}+7\sqrt{2}} \Bigr ) \cr\cr
& & + \frac{1}{3\pi i} \Bigl [ {\rm Li}_2 ((\sqrt{3}-2)e^{i\phi_0}) - {\rm Li}_2 ((\sqrt{3}-2)e^{-i\phi_0}) - {\rm Li}_2 ((2\sqrt{6}-5)e^{i\phi_0}) \cr\cr
& & + {\rm Li}_2 ((2\sqrt{6}-5)e^{-i\phi_0}) + {\rm Li}_2 ((3-2\sqrt{2})e^{i\phi_0}) - {\rm Li}_2 ((3-2\sqrt{2})e^{-i\phi_0}) - {\rm Li}_2 (e^{i\phi_0}) \cr\cr
& & + {\rm Li}_2 (e^{-i\phi_0}) \Bigr ] \cr\cr
& = & 1.127778363805542...
\eeqs
where $\phi_0 = \tan^{-1} (-2\sqrt{6}) = 1.772154247585227...$.

\subsubsection{Net 15}

Similar to the Kagom\'e lattice, $(3.6.3.6)$, net 15 is a simple combination of hexagons and triangles as shown in Fig. \ref{netfig1} (d).
A primitive unit cell contains eleven vertices $\nu_{15}=11$. There are two $6^3$, four $3^2.6^2$ and five $3.6.3.6$ vertices in each unit cell, so that the effective coordination number is $\kappa_{15}=42/11$. Referring to Fig. \ref{netfig1} (d), if one takes the direction to the right as $\theta_1$ and the direction to the upper-right as $\theta_2$, then
\beq
M_{15}(\theta_1,\theta_2) = \left( \begin{array}{ccccccccccc}
4 & -1 & 0 & 0 & 0 & -\alpha & -\alpha & 0 & 0 & -1 & 0 \\
-1 & 4 & -1 & 0 & 0 & -\alpha & 0 & 0 & 0 & 0 & -1 \\
0 & -1 & 4 & -1 & 0 & 0 & 0 & 0 & -\beta & 0 & -1 \\
0 & 0 & -1 & 4 & -1 & 0 & 0 & -\beta & -\beta & 0 & 0 \\
0 & 0 & 0 & -1 & 3 & -1 & 0 & 0 & 0 & -\frac{\beta}{\alpha} & 0 \\
-\frac{1}{\alpha} & -\frac{1}{\alpha} & 0 & 0 & -1 & 4 & -1 & 0 & 0 & 0 & 0 \\
-\frac{1}{\alpha} & 0 & 0 & 0 & 0 & -1 & 4 & -1 & 0 & 0 & -1 \\
0 & 0 & 0 & -\frac{1}{\beta} & 0 & 0 & -1 & 4 & -1 & 0 & -1 \\
0 & 0 & -\frac{1}{\beta} & -\frac{1}{\beta} & 0 & 0 & 0 & -1 & 4 & -1 & 0 \\
-1 & 0 & 0 & 0 & -\frac{\alpha}{\beta} & 0 & 0 & 0 & -1 & 3 & 0 \\
0 & -1 & -1 & 0 & 0 & 0 & -1 & -1 & 0 & 0 & 4
\end{array} \right ) \ ,
\eeq
with determinant
\beqs
D_{15}(\theta_1,\theta_2)
& = & 4 \{ 39910 - 16905 (\cos \theta_1 + \cos \theta_2) - 1270 \cos (\theta_1 - \theta_2) + 356 (\cos^2 \theta_1 + \cos^2 \theta_2) \cr\cr
& & - 5352 \cos \theta_1 \cos \theta_2 - 25 (\cos \theta_1 + \cos \theta_2) \cos (\theta_1 - \theta_2) \cr\cr 
& & - 70 \cos \theta_1 \cos \theta_2 (\cos \theta_1 + \cos \theta_2) \} \ .
\eeqs
However, if one still takes the direction to the right as $\theta_1$ but the direction to the upper-left as $\theta_2$, then
\beq
\bar M_{15}(\theta_1,\theta_2) = \left( \begin{array}{ccccccccccc}
4 & -1 & 0 & 0 & 0 & -\alpha & -\alpha & 0 & 0 & -1 & 0 \\
-1 & 4 & -1 & 0 & 0 & -\alpha & 0 & 0 & 0 & 0 & -1 \\
0 & -1 & 4 & -1 & 0 & 0 & 0 & 0 & -\alpha\beta & 0 & -1 \\
0 & 0 & -1 & 4 & -1 & 0 & 0 & -\alpha\beta & -\alpha\beta & 0 & 0 \\
0 & 0 & 0 & -1 & 3 & -1 & 0 & 0 & 0 & -\beta & 0 \\
-\frac{1}{\alpha} & -\frac{1}{\alpha} & 0 & 0 & -1 & 4 & -1 & 0 & 0 & 0 & 0 \\
-\frac{1}{\alpha} & 0 & 0 & 0 & 0 & -1 & 4 & -1 & 0 & 0 & -1 \\
0 & 0 & 0 & -\frac{1}{\alpha\beta} & 0 & 0 & -1 & 4 & -1 & 0 & -1 \\
0 & 0 & -\frac{1}{\alpha\beta} & -\frac{1}{\alpha\beta} & 0 & 0 &, 0 & -1 & 4 & -1 & 0 \\
-1 & 0 & 0 & 0 & -\frac{1}{\beta} & 0 & 0 & 0 & -1 & 3 & 0 \\
0 & -1 & -1 & 0 & 0 & 0 & -1 & -1 & 0 & 0 & 4
\end{array} \right ) \ ,
\eeq
with determinant
\beqs
\bar D_{15}(\theta_1,\theta_2)
& = & 4\{ 40266 - 16975\cos \theta_1 - 1270 \cos \theta_2 - 16905\cos (\theta_1 + \theta_2) - 356 \cos^2 \theta_2 \cr\cr
& & - 25 \cos \theta_1 \cos \theta_2 - 5352 \cos \theta_1 \cos (\theta_1 + \theta_2) - 25 \cos \theta_2 \cos (\theta_1 + \theta_2) \cr\cr
& & + 70 \cos^3 \theta_1 + 70 \cos \theta_1 \cos^2 \theta_2 - 70 \cos^2 \theta_1 \cos (\theta_1 + \theta_2) \cr\cr
& & + 712 \cos \theta_1 \cos \theta_2 \cos (\theta_1 + \theta_2) - 140 \cos^2 \theta_1 \cos \theta_2 \cos (\theta_1 + \theta_2) \} \ ,
\eeqs
which looks distinct from $D_{15}(\theta_1,\theta_2)$.
Nevertheless, both determinants give the same asymptotic growth constant and we have the integral identity
\beqs
z_{15} & = & \frac{1}{11} \int_{-\pi}^{\pi} \frac{d\theta_1}{2\pi} \int_{-\pi}^{\pi} \frac{d\theta_2}{2\pi}
\ln \Big [ D_{15}(\theta_1,\theta_2) \Big ] \cr\cr
& = & \frac{1}{11} \int_{-\pi}^{\pi} \frac{d\theta_1}{2\pi} \int_{-\pi}^{\pi} \frac{d\theta_2}{2\pi}
\ln \Big [ \bar D_{15}(\theta_1,\theta_2) \Big ] = 1.073270254423056... 
\eeqs
In contrast to the integral identities in \cite{sti} which was obtained by different choice of unit cells, here we obtain an integral identity by different choice of directions in the calculation.

\subsubsection{Net 16}

Net 16 is a combination of triangles, squares and hexagons.
There are two kinds of net 16, namely net 16(a) and net 16(b). Their unit cells are shown in Fig. \ref{netfig1} (e) and (f), respectively, and both of them contain twelve vertices $\nu_{16(a)}=\nu_{16(b)}=12$. For net 16(a), there are six $3^2.4.3.4$ and six $3.4.6.4$ vertices in each unit cell, so that the effective coordination number is $\kappa_{16(a)}=9/2$. We have
\beq
M_{16(a)}(\theta_1,\theta_2) = \left( \begin{array}{cccccccccccc}
5 & -1 & 0 & -\frac{1}{\alpha\beta} & 0 & 0 & -\frac{1}{\beta} & 0 & -1 & -1 & 0 & 0 \\
-1 & 4 & -1 & 0 & 0 & -\frac{1}{\beta} & 0 & 0 & 0 & -1 & 0 & 0 \\
0 & -1 & 4 & -1 & 0 & 0 & 0 & -\alpha & 0 & 0 & -1 & 0 \\
-\alpha\beta & 0 & -1 & 5 & -1 & 0 & -\alpha & 0 & 0 & 0 & -1 & 0 \\
0 & 0 & 0 & -1 & 4 & -1 & 0 & 0 & -\alpha\beta & 0 & -1 & 0 \\
0 & -\beta & 0 & 0 & -1 & 4 & -1 & 0 & 0 & 0 & 0 & -1 \\
-\beta & 0 & 0 & -\frac{1}{\alpha} & 0 & -1 & 5 & -1 & 0 & 0 & 0 & -1 \\
0 & 0 & -\frac{1}{\alpha} & 0 & 0 & 0 & -1 & 4 & -1 & 0 & 0 & -1 \\
-1 & 0 & 0 & 0 & -\frac{1}{\alpha\beta} & 0 & 0 & -1 & 4 & -1 & 0 & 0 \\
-1 & -1 & 0 & 0 & 0 & 0 & 0 & 0 & -1 & 5 & -1 & -1 \\
0 & 0 & -1 & -1 & -1 & 0 & 0 & 0 & 0 & -1 & 5 & -1 \\
0 & 0 & 0 & 0 & 0 & -1 & -1 & -1 & 0 & -1 & -1 & 5
\end{array} \right ) \ . 
\eeq
The determinant can be calculated to be
\beqs
D_{16(a)}(\theta_1,\theta_2)
& = & 112 \{ 49746 - 15900 (\cos \theta_1 + \cos \theta_2) - 15899 \cos (\theta_1 + \theta_2) - 749 \cos \theta_1 \cos \theta_2 \cr\cr
& & - 749 (\cos \theta_1 + \cos \theta_2) \cos (\theta_1 + \theta_2) + (\cos^3 \theta_1 + \cos^3 \theta_2) \cr\cr
& & - (\cos^2 \theta_1 + \cos^2 \theta_2) \cos (\theta_1 + \theta_2) + 204 \cos \theta_1 \cos \theta_2 \cos (\theta_1 + \theta_2) \cr\cr
& & - 2 \cos \theta_1 \cos \theta_2 (\cos \theta_1 + \cos \theta_2) \cos (\theta_1 + \theta_2) \} \ ,
\eeqs
such that
\beq
z_{16(a)} = \frac{1}{12} \int_{-\pi}^{\pi} \frac{d\theta_1}{2\pi} \int_{-\pi}^{\pi} \frac{d\theta_2}{2\pi}
\ln \Big [ D_{16(a)}(\theta_1,\theta_2) \Big ] = 1.280287248642483...  
\eeq

For net 16(b), there are six $3^3.4^2$ and six $3.4.6.4$ vertices in each unit cell, so that the effective coordination number is also $\kappa_{16(b)}=9/2$. We have
\beq
M_{16(b)}(\theta_1,\theta_2) = \left( \begin{array}{cccccccccccc}
4 & -1 & 0 & -\alpha & 0 & 0 & 0 & -\beta & 0 & -1 & 0 & 0 \\
-1 & 5 & -1 & 0 & 0 & 0 & -\beta & 0 & 0 & -1 & -1 & 0 \\
0 & -1 & 4 & -1 & 0 & -\beta & 0 & 0 & 0 & 0 & -1 & 0 \\
-\frac{1}{\alpha} & 0 & -1 & 4 & -1 & 0 & 0 & 0 & 0 & 0 & 0 & -1 \\
0 & 0 & 0 & -1 & 5 & -1 & -1 & 0 & 0 & -\frac{1}{\alpha} & 0 & -1 \\
0 & 0 & -\frac{1}{\beta} & 0 & -1 & 4 & -1 & 0 & -\frac{1}{\alpha} & 0 & 0 & 0 \\
0 & -\frac{1}{\beta} & 0 & 0 & -1 & -1 & 5 & -1 & 0 & 0 & 0 & -1 \\
-\frac{1}{\beta} & 0 & 0 & 0 & 0 & 0 & -1 & 4 & -1 & 0 & 0 & -1 \\
0 & 0 & 0 & 0 & 0 & -\alpha & 0 & -1 & 4 & -1 & -1 & 0 \\
-1 & -1 & 0 & 0 & -\alpha & 0 & 0 & 0 & -1 & 5 & -1 & 0 \\
0 & -1 & -1 & 0 & 0 & 0 & 0 & 0 & -1 & -1 & 5 & -1 \\
0 & 0 & 0 & -1 & -1 & 0 & -1 & -1 & 0 & 0 & -1 & 5
\end{array} \right ) \ . 
\eeq
The determinant can be calculated to be
\beqs
D_{16(b)}(\theta_1,\theta_2)
& = & 16 \{ 337746 - 109263 (\cos \theta_1 + \cos \theta_2) - 109262 \cos (\theta_1 - \theta_2) \cr\cr
& & - 4024 \cos \theta_1 \cos \theta_2 - 4023 (\cos \theta_1 + \cos \theta_2) \cos (\theta_1 - \theta_2) + (\cos^3 \theta_1 + \cos^3 \theta_2) \cr\cr 
& & + 2118 \cos \theta_1 \cos \theta_2 \cos (\theta_1 - \theta_2) - \sin \theta_1 \sin \theta_2 (\cos^2 \theta_1 + \cos^2 \theta_2) \cr\cr
& & - 3 \cos \theta_1 \cos \theta_2 \cos (\theta_1 - \theta_2) (\cos \theta_1 + \cos \theta_2) - 2 \cos^2 \theta_1 \cos^2 \theta_2 \cos (\theta_1 - \theta_2) \} \ , \cr & &
\eeqs
such that
\beq
z_{16(b)} = \frac{1}{12} \int_{-\pi}^{\pi} \frac{d\theta_1}{2\pi} \int_{-\pi}^{\pi} \frac{d\theta_2}{2\pi}
\ln \Big [ D_{16(b)}(\theta_1,\theta_2) \Big ] = 1.277617926708331...  
\eeq
which is slightly less than $z_{16(a)}$.

\subsubsection{Net 17}

Net 17 is obtained by intergrowth of $3^2.4.3.4$ and $3.4.6.4$ vertices. A primitive unit cell containing twenty-eight vertices $\nu_{17}=28$ is shown in Fig. \ref{netfig1} (g). There are sixteen $3^2.4.3.4$ vertices and twelve $3.4.6.4$ in each unit cell, so that the effective coordination number is $\kappa_{17}=32/7$. Let us write $M_{17}(\theta_1,\theta_2) = M_{17}^\prime - M_{17}^{\prime\prime}(\theta_1,\theta_2)$, where $M_{17}^\prime$ is a diagonal matrix with diagonal elements $\{5,5,5,5,5,5,5,5,5,5,5,5,4,4,4,4,4,4,5,5,4,4,4,4,4,4,5,5\}$ and 
\beqs
\lefteqn{M_{17}^{\prime\prime}(\theta_1,\theta_2)} \cr\cr
& = & \left( \begin{array}{cccccccccccccccccccccccccccc}
0 & 1 & 0 & 0 & 0 & 0 & 0 & 0 & 0 & 0 & 0 & 0 & 0 & 0 & \alpha & 0 & 0 & 0 & 0 & 1 & 0 & 0 & 0 & 0 & 0 & 0 & 0 & 0 \\
1 & 0 & 1 & 0 & 0 & 0 & 0 & 0 & 0 & 0 & 0 & 0 & 0 & \alpha & 0 & 0 & 0 & 0 & 0 & 1 & 1 & 0 & 0 & 0 & 0 & 0 & 0 & 0 \\
0 & 1 & 0 & 1 & 0 & 0 & 0 & 0 & 0 & 0 & 0 & \alpha & 0 & \alpha & 0 & 0 & 0 & 0 & 0 & 0 & 0 & 1 & 0 & 0 & 0 & 0 & 0 & 0 \\
0 & 0 & 1 & 0 & 1 & 0 & 0 & 0 & 0 & \alpha & 0 & \alpha & 0 & 0 & 0 & 0 & 0 & 0 & 0 & 0 & 0 & 1 & 0 & 0 & 0 & 0 & 0 & 0 \\
0 & 0 & 0 & 1 & 0 & 1 & 0 & 0 & 0 & \alpha & 0 & 0 & 0 & 0 & 0 & \alpha\beta & 0 & 0 & 0 & 0 & 0 & 0 & 1 & 0 & 0 & 0 & 0 & 0 \\
0 & 0 & 0 & 0 & 1 & 0 & 1 & 0 & 0 & 0 & 0 & 0 & 0 & 0 & \alpha\beta & 0 & 0 & 0 & 0 & 0 & \beta & 0 & 1 & 0 & 0 & 0 & 0 & 0 \\
0 & 0 & 0 & 0 & 0 & 1 & 0 & 1 & 0 & 0 & 0 & 0 & 0 & 0 & 0 & 0 & 0 & 0 & \beta & 0 & \beta & 0 & 0 & 1 & 0 & 0 & 0 & 0 \\
0 & 0 & 0 & 0 & 0 & 0 & 1 & 0 & 1 & 0 & 0 & 0 & 0 & 0 & 0 & 0 & \beta & 0 & \beta & 0 & 0 & 0 & 0 & 1 & 0 & 0 & 0 & 0 \\
0 & 0 & 0 & 0 & 0 & 0 & 0 & 1 & 0 & 1 & 1 & 0 & 0 & 0 & 0 & 0 & \beta & 0 & 0 & 0 & 0 & 0 & 0 & 0 & 1 & 0 & 0 & 0 \\
0 & 0 & 0 & \frac{1}{\alpha} & \frac{1}{\alpha} & 0 & 0 & 0 & 1 & 0 & 1 & 0 & 0 & 0 & 0 & \beta & 0 & 0 & 0 & 0 & 0 & 0 & 0 & 0 & 0 & 0 & 0 & 0 \\
0 & 0 & 0 & 0 & 0 & 0 & 0 & 0 & 1 & 1 & 0 & 1 & 0 & 0 & 0 & 0 & 0 & 0 & 0 & 0 & 0 & 0 & 0 & 0 & 1 & 0 & 1 & 0 \\
0 & 0 & \frac{1}{\alpha} & \frac{1}{\alpha} & 0 & 0 & 0 & 0 & 0 & 0 & 1 & 0 & 1 & 0 & 0 & 0 & 0 & 0 & 0 & 0 & 0 & 0 & 0 & 0 & 0 & 0 & 1 & 0 \\
0 & 0 & 0 & 0 & 0 & 0 & 0 & 0 & 0 & 0 & 0 & 1 & 0 & 1 & 0 & 0 & 0 & 1 & 0 & 0 & 0 & 0 & 0 & 0 & 0 & 0 & 1 & 0 \\
0 & \frac{1}{\alpha} & \frac{1}{\alpha} & 0 & 0 & 0 & 0 & 0 & 0 & 0 & 0 & 0 & 1 & 0 & 1 & 0 & 0 & 0 & 0 & 0 & 0 & 0 & 0 & 0 & 0 & 0 & 0 & 0 \\
\frac{1}{\alpha} & 0 & 0 & 0 & 0 & \frac{1}{\alpha\beta} & 0 & 0 & 0 & 0 & 0 & 0 & 0 & 1 & 0 & 1 & 0 & 0 & 0 & 0 & 0 & 0 & 0 & 0 & 0 & 0 & 0 & 0 \\
0 & 0 & 0 & 0 & \frac{1}{\alpha\beta} & 0 & 0 & 0 & 0 & \frac{1}{\beta} & 0 & 0 & 0 & 0 & 1 & 0 & 1 & 0 & 0 & 0 & 0 & 0 & 0 & 0 & 0 & 0 & 0 & 0 \\
0 & 0 & 0 & 0 & 0 & 0 & 0 & \frac{1}{\beta} & \frac{1}{\beta} & 0 & 0 & 0 & 0 & 0 & 0 & 1 & 0 & 1 & 0 & 0 & 0 & 0 & 0 & 0 & 0 & 0 & 0 & 0 \\
0 & 0 & 0 & 0 & 0 & 0 & 0 & 0 & 0 & 0 & 0 & 0 & 1 & 0 & 0 & 0 & 1 & 0 & 1 & 0 & 0 & 0 & 0 & 0 & 0 & 0 & 0 & 1 \\
0 & 0 & 0 & 0 & 0 & 0 & \frac{1}{\beta} & \frac{1}{\beta} & 0 & 0 & 0 & 0 & 0 & 0 & 0 & 0 & 0 & 1 & 0 & 1 & 0 & 0 & 0 & 0 & 0 & 0 & 0 & 1 \\
1 & 1 & 0 & 0 & 0 & 0 & 0 & 0 & 0 & 0 & 0 & 0 & 0 & 0 & 0 & 0 & 0 & 0 & 1 & 0 & 1 & 0 & 0 & 0 & 0 & 0 & 0 & 1 \\
0 & 1 & 0 & 0 & 0 & \frac{1}{\beta} & \frac{1}{\beta} & 0 & 0 & 0 & 0 & 0 & 0 & 0 & 0 & 0 & 0 & 0 & 0 & 1 & 0 & 1 & 0 & 0 & 0 & 1 & 0 & 0 \\
0 & 0 & 1 & 1 & 0 & 0 & 0 & 0 & 0 & 0 & 0 & 0 & 0 & 0 & 0 & 0 & 0 & 0 & 0 & 0 & 1 & 0 & 1 & 0 & 0 & 0 & 0 & 0 \\
0 & 0 & 0 & 0 & 1 & 1 & 0 & 0 & 0 & 0 & 0 & 0 & 0 & 0 & 0 & 0 & 0 & 0 & 0 & 0 & 0 & 1 & 0 & 1 & 0 & 0 & 0 & 0 \\
0 & 0 & 0 & 0 & 0 & 0 & 1 & 1 & 0 & 0 & 0 & 0 & 0 & 0 & 0 & 0 & 0 & 0 & 0 & 0 & 0 & 0 & 1 & 0 & 1 & 0 & 0 & 0 \\
0 & 0 & 0 & 0 & 0 & 0 & 0 & 0 & 1 & 0 & 1 & 0 & 0 & 0 & 0 & 0 & 0 & 0 & 0 & 0 & 0 & 0 & 0 & 1 & 0 & 1 & 0 & 0 \\
0 & 0 & 0 & 0 & 0 & 0 & 0 & 0 & 0 & 0 & 0 & 0 & 0 & 0 & 0 & 0 & 0 & 0 & 0 & 0 & 1 & 0 & 0 & 0 & 1 & 0 & 1 & 1 \\
0 & 0 & 0 & 0 & 0 & 0 & 0 & 0 & 0 & 0 & 1 & 1 & 1 & 0 & 0 & 0 & 0 & 0 & 0 & 0 & 0 & 0 & 0 & 0 & 0 & 1 & 0 & 1 \\
0 & 0 & 0 & 0 & 0 & 0 & 0 & 0 & 0 & 0 & 0 & 0 & 0 & 0 & 0 & 0 & 0 & 1 & 1 & 1 & 0 & 0 & 0 & 0 & 0 & 1 & 1 & 0
\end{array} \right ) \ . \cr & &
\eeqs
The determinant can be calculated to be
\beqs
D_{17}(\theta_1,\theta_2)
& = & 16 \{ 466754878482464 - 193533953205944 (\cos \theta_1 + \cos \theta_2) \cr\cr
& & + 3411923764300 (\cos^2 \theta_1 + \cos^2 \theta_2) - 84153824536936 \cos \theta_1 \cos \theta_2 \cr\cr
& & - 1683339412 (\cos^3 \theta_1 + \cos^3 \theta_2) - 1185500641948 \cos \theta_1 \cos \theta_2 (\cos \theta_1 + \cos \theta_2) \cr\cr 
& & + 38809 (\cos^4 \theta_1 + \cos^4 \theta_2) - 729383812 \cos \theta_1 \cos \theta_2 (\cos^2 \theta_1 + \cos^2 \theta_2) \cr\cr 
& & + 18854974886 \cos^2 \theta_1 \cos^2 \theta_2 - 11032 \cos \theta_1 \cos \theta_2 (\cos^3 \theta_1 + \cos^3 \theta_2) \cr\cr
& & - 11681168 \cos^2 \theta_1 \cos^2 \theta_2 (\cos \theta_1 + \cos \theta_2) \cr\cr 
& & + 784 \cos^2 \theta_1 \cos^2 \theta_2 (\cos^2 \theta_1 + \cos^2 \theta_2) - 1568 \cos^3 \theta_1 \cos^3 \theta_2 \} \ ,
\eeqs
such that
\beq
z_{17} = \frac{1}{28} \int_{-\pi}^{\pi} \frac{d\theta_1}{2\pi} \int_{-\pi}^{\pi} \frac{d\theta_2}{2\pi}
\ln \Big [ D_{17}(\theta_1,\theta_2) \Big ] = 1.299177753544099...  
\eeq

\subsection{Nets with pentagons}

In this subsection, we consider important nets which involve pentagons.
Most of them also contain triangles and squares, in addition to pentagons.

\subsubsection{Net 18}

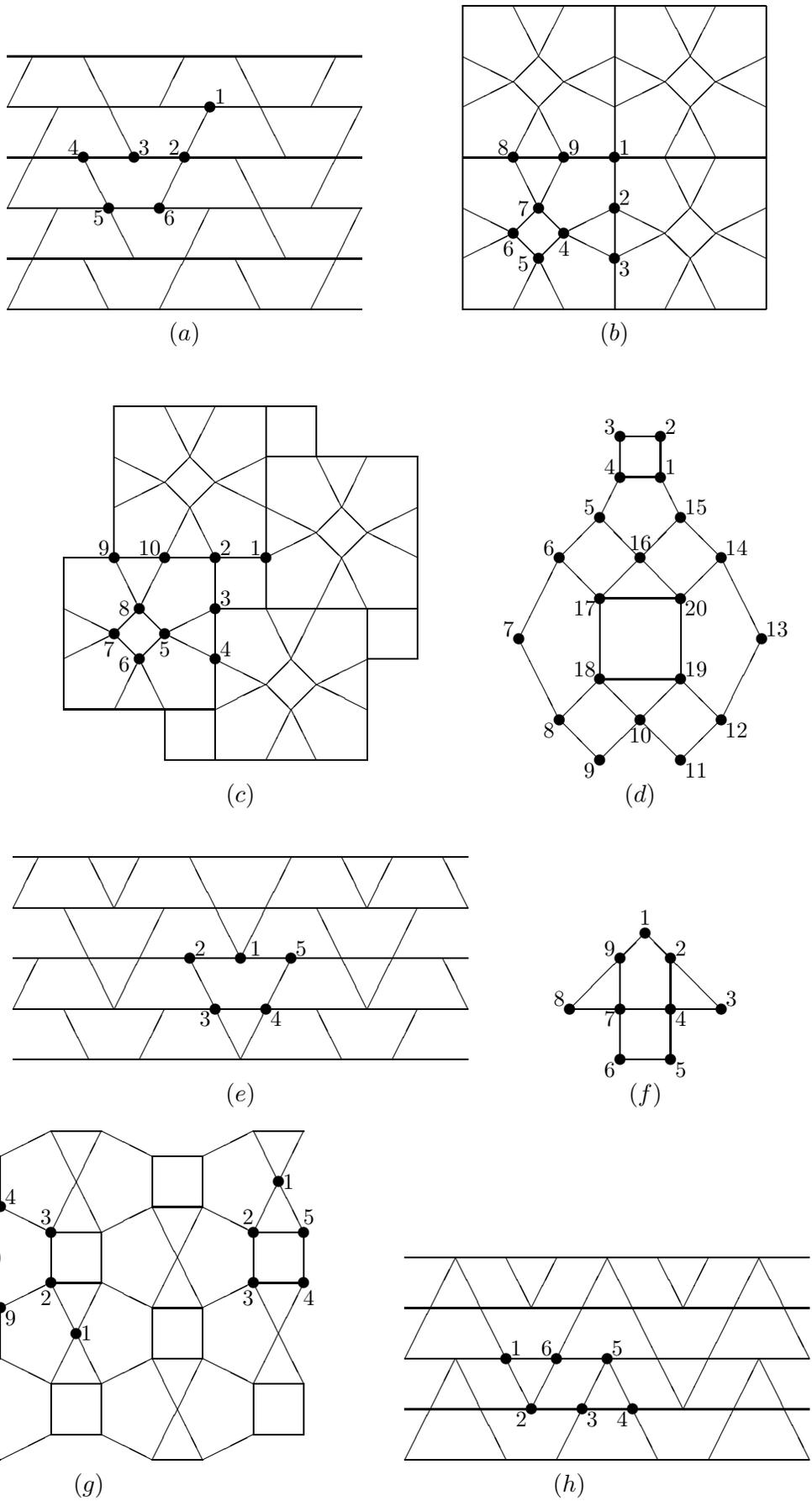
\begin{figure}[htbp]
\unitlength 0.8mm \hspace*{5mm}
\begin{picture}(150,60)
\multiput(0,0)(0,10){6}{\line(1,0){70}}
\multiput(0,0)(30,0){3}{\line(1,2){10}}
\multiput(0,20)(30,0){3}{\line(1,2){10}}
\multiput(0,40)(30,0){3}{\line(1,2){5}}
\multiput(20,0)(30,0){2}{\line(-1,2){5}}
\multiput(25,10)(30,0){2}{\line(-1,2){10}}
\multiput(25,30)(30,0){2}{\line(-1,2){10}}
\multiput(20,20)(10,0){2}{\circle*{2}}
\multiput(15,30)(10,0){3}{\circle*{2}}
\put(40,40){\circle*{2}}
\put(42,42){\makebox(0,0){\footnotesize 1}}
\put(33,32){\makebox(0,0){\footnotesize 2}}
\put(27,32){\makebox(0,0){\footnotesize 3}}
\put(13,32){\makebox(0,0){\footnotesize 4}}
\put(18,18){\makebox(0,0){\footnotesize 5}}
\put(32,18){\makebox(0,0){\footnotesize 6}}
\put(35,-5){\makebox(0,0){$(a)$}}

\multiput(90,0)(0,30){3}{\line(1,0){60}}
\multiput(90,0)(30,0){3}{\line(0,1){60}}
\multiput(100,0)(30,0){2}{\line(1,2){5}}
\multiput(110,0)(30,0){2}{\line(-1,2){5}}
\multiput(90,10)(30,0){2}{\line(2,1){10}}
\multiput(90,20)(30,0){2}{\line(2,-1){10}}
\multiput(100,30)(30,0){2}{\line(1,-2){5}}
\multiput(110,30)(30,0){2}{\line(-1,-2){5}}
\multiput(120,10)(30,0){2}{\line(-2,1){10}}
\multiput(120,20)(30,0){2}{\line(-2,-1){10}}
\multiput(100,15)(30,0){2}{\line(1,1){5}}
\multiput(100,15)(30,0){2}{\line(1,-1){5}}
\multiput(110,15)(30,0){2}{\line(-1,1){5}}
\multiput(110,15)(30,0){2}{\line(-1,-1){5}}
\multiput(100,30)(30,0){2}{\line(1,2){5}}
\multiput(110,30)(30,0){2}{\line(-1,2){5}}
\multiput(90,40)(30,0){2}{\line(2,1){10}}
\multiput(90,50)(30,0){2}{\line(2,-1){10}}
\multiput(100,60)(30,0){2}{\line(1,-2){5}}
\multiput(110,60)(30,0){2}{\line(-1,-2){5}}
\multiput(120,40)(30,0){2}{\line(-2,1){10}}
\multiput(120,50)(30,0){2}{\line(-2,-1){10}}
\multiput(100,45)(30,0){2}{\line(1,1){5}}
\multiput(100,45)(30,0){2}{\line(1,-1){5}}
\multiput(110,45)(30,0){2}{\line(-1,1){5}}
\multiput(110,45)(30,0){2}{\line(-1,-1){5}}
\multiput(120,10)(0,10){2}{\circle*{2}}
\multiput(100,30)(10,0){3}{\circle*{2}}
\multiput(100,15)(10,0){2}{\circle*{2}}
\multiput(105,10)(0,10){2}{\circle*{2}}
\put(122,32){\makebox(0,0){\footnotesize 1}}
\put(122,22){\makebox(0,0){\footnotesize 2}}
\put(122,8){\makebox(0,0){\footnotesize 3}}
\put(110,12){\makebox(0,0){\footnotesize 4}}
\put(102,9){\makebox(0,0){\footnotesize 5}}
\put(99,12){\makebox(0,0){\footnotesize 6}}
\put(102,20){\makebox(0,0){\footnotesize 7}}
\put(98,32){\makebox(0,0){\footnotesize 8}}
\put(112,32){\makebox(0,0){\footnotesize 9}}
\put(120,-5){\makebox(0,0){$(b)$}}
\end{picture}

\vspace*{15mm}

\begin{picture}(120,70)
\multiput(20,0)(10,30){2}{\line(1,0){40}}
\put(40,60){\line(1,0){30}}
\put(0,10){\line(1,0){30}}
\put(60,20){\line(1,0){10}}
\multiput(0,40)(10,30){2}{\line(1,0){40}}
\multiput(0,10)(10,30){2}{\line(0,1){30}}
\multiput(20,0)(30,60){2}{\line(0,1){10}}
\put(70,20){\line(0,1){10}}
\multiput(30,0)(10,30){2}{\line(0,1){40}}
\multiput(60,0)(10,30){2}{\line(0,1){30}}
\multiput(10,10)(30,-10){2}{\line(1,2){5}}
\multiput(20,10)(30,-10){2}{\line(-1,2){5}}
\multiput(10,40)(30,-10){2}{\line(1,-2){5}}
\multiput(20,40)(30,-10){2}{\line(-1,-2){5}}
\multiput(0,20)(30,-10){2}{\line(2,1){10}}
\multiput(0,30)(30,-10){2}{\line(2,-1){10}}
\multiput(30,20)(30,-10){2}{\line(-2,1){10}}
\multiput(30,30)(30,-10){2}{\line(-2,-1){10}}
\multiput(15,20)(5,5){2}{\line(-1,1){5}}
\multiput(15,20)(-5,5){2}{\line(1,1){5}}
\multiput(45,10)(5,5){2}{\line(-1,1){5}}
\multiput(45,10)(-5,5){2}{\line(1,1){5}}
\multiput(20,40)(30,-10){2}{\line(1,2){5}}
\multiput(30,40)(30,-10){2}{\line(-1,2){5}}
\multiput(20,70)(30,-10){2}{\line(1,-2){5}}
\multiput(30,70)(30,-10){2}{\line(-1,-2){5}}
\multiput(10,50)(30,-10){2}{\line(2,1){10}}
\multiput(10,60)(30,-10){2}{\line(2,-1){10}}
\multiput(40,50)(30,-10){2}{\line(-2,1){10}}
\multiput(40,60)(30,-10){2}{\line(-2,-1){10}}
\multiput(25,50)(5,5){2}{\line(-1,1){5}}
\multiput(25,50)(-5,5){2}{\line(1,1){5}}
\multiput(55,40)(5,5){2}{\line(-1,1){5}}
\multiput(55,40)(-5,5){2}{\line(1,1){5}}
\multiput(30,20)(0,10){2}{\circle*{2}}
\multiput(10,40)(10,0){4}{\circle*{2}}
\multiput(10,25)(10,0){2}{\circle*{2}}
\multiput(15,20)(0,10){2}{\circle*{2}}
\put(38,42){\makebox(0,0){\footnotesize 1}}
\put(32,42){\makebox(0,0){\footnotesize 2}}
\put(32,32){\makebox(0,0){\footnotesize 3}}
\put(32,22){\makebox(0,0){\footnotesize 4}}
\put(20,22){\makebox(0,0){\footnotesize 5}}
\put(12,19){\makebox(0,0){\footnotesize 6}}
\put(9,22){\makebox(0,0){\footnotesize 7}}
\put(12,30){\makebox(0,0){\footnotesize 8}}
\put(8,42){\makebox(0,0){\footnotesize 9}}
\put(17,42){\makebox(0,0){\footnotesize 10}}
\put(35,-7){\makebox(0,0){$(c)$}}

\multiput(106,16)(0,16){2}{\line(1,0){16}}
\multiput(106,16)(16,0){2}{\line(0,1){16}}
\multiput(106,0)(0,32){2}{\line(1,1){16}}
\multiput(122,0)(0,32){2}{\line(-1,1){16}}
\multiput(98,8)(0,32){2}{\line(1,1){8}}
\multiput(98,8)(0,32){2}{\line(1,-1){8}}
\multiput(130,8)(0,32){2}{\line(-1,1){8}}
\multiput(130,8)(0,32){2}{\line(-1,-1){8}}
\multiput(90,24)(40,-16){2}{\line(1,2){8}}
\multiput(90,24)(40,16){2}{\line(1,-2){8}}
\multiput(110,56)(0,8){2}{\line(1,0){8}}
\multiput(110,56)(8,0){2}{\line(0,1){8}}
\put(106,48){\line(1,2){4}}
\put(122,48){\line(-1,2){4}}
\multiput(106,0)(0,16){4}{\circle*{2}}
\multiput(122,0)(0,16){4}{\circle*{2}}
\multiput(98,8)(0,32){2}{\circle*{2}}
\multiput(130,8)(0,32){2}{\circle*{2}}
\multiput(114,8)(0,32){2}{\circle*{2}}
\multiput(90,24)(48,0){2}{\circle*{2}}
\multiput(110,56)(0,8){2}{\circle*{2}}
\multiput(118,56)(0,8){2}{\circle*{2}}
\put(120,58){\makebox(0,0){\footnotesize 1}}
\put(120,66){\makebox(0,0){\footnotesize 2}}
\put(108,66){\makebox(0,0){\footnotesize 3}}
\put(108,58){\makebox(0,0){\footnotesize 4}}
\put(104,50){\makebox(0,0){\footnotesize 5}}
\put(96,42){\makebox(0,0){\footnotesize 6}}
\put(88,26){\makebox(0,0){\footnotesize 7}}
\put(96,6){\makebox(0,0){\footnotesize 8}}
\put(104,-2){\makebox(0,0){\footnotesize 9}}
\put(114,5){\makebox(0,0){\footnotesize 10}}
\put(125,-2){\makebox(0,0){\footnotesize 11}}
\put(133,6){\makebox(0,0){\footnotesize 12}}
\put(141,26){\makebox(0,0){\footnotesize 13}}
\put(133,42){\makebox(0,0){\footnotesize 14}}
\put(125,50){\makebox(0,0){\footnotesize 15}}
\put(114,43){\makebox(0,0){\footnotesize 16}}
\put(103,30){\makebox(0,0){\footnotesize 17}}
\put(103,18){\makebox(0,0){\footnotesize 18}}
\put(125,18){\makebox(0,0){\footnotesize 19}}
\put(125,30){\makebox(0,0){\footnotesize 20}}
\put(114,-7){\makebox(0,0){$(d)$}}
\end{picture}

\vspace*{15mm}

\begin{picture}(140,40)
\multiput(0,0)(0,10){5}{\line(1,0){90}}
\multiput(15,0)(50,0){2}{\line(-1,2){5}}
\multiput(25,0)(50,0){2}{\line(1,2){5}}
\multiput(45,0)(0,20){2}{\line(-1,2){10}}
\multiput(45,0)(0,20){2}{\line(1,2){10}}
\multiput(20,10)(50,0){2}{\line(-1,2){10}}
\multiput(20,10)(50,0){2}{\line(1,2){10}}
\multiput(20,30)(50,0){2}{\line(-1,2){5}}
\multiput(20,30)(50,0){2}{\line(1,2){5}}
\multiput(0,10)(0,20){2}{\line(1,2){5}}
\multiput(90,10)(0,20){2}{\line(-1,2){5}}
\multiput(40,10)(10,0){2}{\circle*{2}}
\multiput(35,20)(10,0){3}{\circle*{2}}
\put(48,22){\makebox(0,0){\footnotesize 1}}
\put(37,22){\makebox(0,0){\footnotesize 2}}
\put(38,8){\makebox(0,0){\footnotesize 3}}
\put(52,8){\makebox(0,0){\footnotesize 4}}
\put(57,22){\makebox(0,0){\footnotesize 5}}
\put(45,-7){\makebox(0,0){$(e)$}}

\multiput(120,0)(10,0){2}{\line(0,1){20}}
\put(120,0){\line(1,0){10}}
\put(110,10){\line(1,0){30}}
\put(110,10){\line(1,1){15}}
\put(140,10){\line(-1,1){15}}
\multiput(120,0)(0,10){3}{\circle*{2}}
\multiput(130,0)(0,10){3}{\circle*{2}}
\multiput(110,10)(30,0){2}{\circle*{2}}
\put(125,25){\circle*{2}}
\put(125,28){\makebox(0,0){\footnotesize 1}}
\put(132,22){\makebox(0,0){\footnotesize 2}}
\put(142,12){\makebox(0,0){\footnotesize 3}}
\put(132,8){\makebox(0,0){\footnotesize 4}}
\put(132,-2){\makebox(0,0){\footnotesize 5}}
\put(118,-2){\makebox(0,0){\footnotesize 6}}
\put(118,8){\makebox(0,0){\footnotesize 7}}
\put(108,12){\makebox(0,0){\footnotesize 8}}
\put(118,22){\makebox(0,0){\footnotesize 9}}
\put(125,-7){\makebox(0,0){$(f)$}}
\end{picture}

\vspace*{15mm}

\begin{picture}(195,60)
\multiput(15,0)(0,20){2}{\line(1,0){10}}
\multiput(5,5)(30,0){2}{\line(0,1){10}}
\multiput(5,5)(20,15){2}{\line(2,-1){10}}
\multiput(5,15)(20,-15){2}{\line(2,1){10}}
\multiput(15,20)(10,0){2}{\line(0,1){10}}
\multiput(35,5)(0,10){2}{\line(1,0){10}}
\multiput(15,0)(20,15){2}{\line(1,2){10}}
\multiput(15,20)(20,15){2}{\line(1,-2){10}}
\multiput(55,0)(0,20){2}{\line(1,0){10}}
\multiput(45,5)(30,0){2}{\line(0,1){10}}
\multiput(45,5)(20,15){2}{\line(2,-1){10}}
\multiput(45,15)(20,-15){2}{\line(2,1){10}}
\multiput(55,20)(10,0){2}{\line(0,1){10}}
\multiput(75,5)(0,10){2}{\line(1,0){10}}
\multiput(55,0)(20,15){2}{\line(1,2){10}}
\multiput(55,20)(20,15){2}{\line(1,-2){10}}
\multiput(15,30)(0,20){2}{\line(1,0){10}}
\multiput(5,35)(30,0){2}{\line(0,1){10}}
\multiput(5,35)(20,15){2}{\line(2,-1){10}}
\multiput(5,45)(20,-15){2}{\line(2,1){10}}
\multiput(15,50)(10,0){2}{\line(0,1){10}}
\multiput(35,35)(0,10){2}{\line(1,0){10}}
\multiput(15,30)(20,15){2}{\line(1,2){10}}
\multiput(15,50)(20,15){2}{\line(1,-2){10}}
\multiput(55,30)(0,20){2}{\line(1,0){10}}
\multiput(45,35)(30,0){2}{\line(0,1){10}}
\multiput(45,35)(20,15){2}{\line(2,-1){10}}
\multiput(45,45)(20,-15){2}{\line(2,1){10}}
\multiput(55,50)(10,0){2}{\line(0,1){10}}
\multiput(75,35)(0,10){2}{\line(1,0){10}}
\multiput(55,30)(20,15){2}{\line(1,2){10}}
\multiput(55,50)(20,15){2}{\line(1,-2){10}}
\multiput(15,60)(40,0){2}{\line(1,0){10}}
\multiput(25,60)(40,0){2}{\line(2,1){10}}
\multiput(35,65)(40,0){2}{\line(1,0){10}}
\multiput(5,65)(40,0){2}{\line(2,-1){10}}
\multiput(5,15)(0,30){2}{\line(-1,2){5}}
\multiput(5,35)(0,30){2}{\line(-1,-2){5}}
\multiput(85,5)(0,30){2}{\line(0,1){10}}
\multiput(5,35)(0,10){2}{\circle*{2}}
\multiput(15,30)(0,20){2}{\circle*{2}}
\multiput(25,30)(0,20){2}{\circle*{2}}
\multiput(35,35)(0,10){2}{\circle*{2}}
\multiput(20,40)(20,-15){2}{\circle*{2}}
\put(42,25){\makebox(0,0){\footnotesize 1}}
\put(34,32){\makebox(0,0){\footnotesize 2}}
\put(34,48){\makebox(0,0){\footnotesize 3}}
\put(27,52){\makebox(0,0){\footnotesize 4}}
\put(13,52){\makebox(0,0){\footnotesize 5}}
\put(6,48){\makebox(0,0){\footnotesize 6}}
\put(6,32){\makebox(0,0){\footnotesize 7}}
\put(13,28){\makebox(0,0){\footnotesize 8}}
\put(27,28){\makebox(0,0){\footnotesize 9}}
\put(23,40){\makebox(0,0){\footnotesize 10}}
\multiput(75,35)(0,10){2}{\circle*{2}}
\multiput(85,35)(0,10){2}{\circle*{2}}
\put(80,55){\circle*{2}}
\put(82,55){\makebox(0,0){\footnotesize 1}}
\put(74,48){\makebox(0,0){\footnotesize 2}}
\put(74,32){\makebox(0,0){\footnotesize 3}}
\put(86,32){\makebox(0,0){\footnotesize 4}}
\put(86,48){\makebox(0,0){\footnotesize 5}}
\put(42.5,-5){\makebox(0,0){$(g)$}}

\multiput(105,0)(0,10){5}{\line(1,0){80}}
\multiput(115,20)(30,0){3}{\line(1,-2){10}}
\multiput(115,40)(30,0){2}{\line(1,-2){15}}
\put(175,40){\line(1,-2){10}}
\multiput(125,40)(30,0){2}{\line(1,-2){5}}
\multiput(105,0)(30,0){3}{\line(1,2){10}}
\put(105,20){\line(1,2){10}}
\multiput(130,10)(30,0){2}{\line(1,2){15}}
\multiput(130,30)(30,0){2}{\line(1,2){5}}
\multiput(130,10)(10,0){3}{\circle*{2}}
\multiput(125,20)(10,0){3}{\circle*{2}}
\put(127,22){\makebox(0,0){\footnotesize 1}}
\put(128,8){\makebox(0,0){\footnotesize 2}}
\put(142,8){\makebox(0,0){\footnotesize 3}}
\put(148,8){\makebox(0,0){\footnotesize 4}}
\put(147,22){\makebox(0,0){\footnotesize 5}}
\put(133,22){\makebox(0,0){\footnotesize 6}}
\put(137.5,-5){\makebox(0,0){$(h)$}}
\end{picture}

\vspace*{5mm}
\caption{\footnotesize{(a) Net 18. (b) Net 19. (c) Net 20. (d) A unit cell of net 21. (e) Net 22. (f) A unit cell of net 23. (g) Two unit cells of net 24. (h) Net 25. Vertices within a unit cell are labeled.}} 
\label{netfig2}
\end{figure}

Net 18 is a pentagon-only net, and is the dual of the $(3^2.4.3.4)$ lattice. The normal appearance that it contains equal-sided (but not regular) pentagons with two angles of $\pi/2$ is not crucial for the calculation of spanning trees. Let us draw it as shown in Fig. \ref{netfig2} (a) with six vertices in a unit cell $\nu_{18}=6$. There are two $5^4$ and four $5^3$ vertices in each unit cell, so that the effective coordination number is $\kappa_{18}=10/3$. We have
\beq
M_{18}(\theta_1,\theta_2) = \left( \begin{array}{cccccc}
3 & -1 & 0 & 0 & -e^{i(\theta_1+\theta_2)} & -e^{i\theta_2} \\ 
-1 & 4 & -1 & -e^{i\theta_1} & 0 & -1 \\
0 & -1 & 3 & -1 & -e^{i\theta_2} & 0 \\
0 & -e^{-i\theta_1} & -1 & 3 & -1 & 0 \\
-e^{-i(\theta_1+\theta_2)} & 0 & -e^{-i\theta_2} & -1 & 4 & -1 \\
-e^{-i\theta_2} & -1 & 0 & 0 & -1 & 3
\end{array} \right ) \ .
\eeq
The determinant can be calculated to be
\beq
D_{18}(\theta_1,\theta_2) = 4 \{ 84 - 36 (\cos \theta_1 + \cos \theta_2) + (\cos^2 \theta_1 + \cos^2 \theta_2) - 14 \cos \theta_1 \cos \theta_2 \} \ ,
\eeq
which is the same as that for the $(3^2.4.3.4)$ lattice as expected. According to Eq. (\ref{zdual}) for $k=5$, we get
\beq
z_{18} = \frac{2}{3} z_{(3^2.4.3.4)} = \frac{1}{6} \int_{-\pi}^{\pi} \frac{d\theta_1}{2\pi} \int_{-\pi}^{\pi} \frac{d\theta_2}{2\pi}
\ln \Big [ D_{18}(\theta_1,\theta_2) \Big ] = 0.9405704304962232...  
\label{z18}
\eeq
where $z_{(3^2.4.3.4)}$ is given in \cite{sti}. A closed-form expression for the integral in Eq. (\ref{z18}) is given in \cite{lamb} such that
\beqs
z_{18} & = & \frac{\ln 6}{3} + \frac{4C}{3\pi} + \frac{8}{3\pi} \Big [ {\rm Ti}_2(pq,q) + {\rm Ti}_2(pq,-q) + {\rm Ti}_2 \Big (\frac{p}{q},\frac{1}{q} \Big ) + {\rm Ti}_2 \Big (\frac{p}{q},-\frac{1}{q} \Big ) \cr\cr
& & - {\rm Ti}_2 \Big (pq,\frac{q}{p} \Big ) - {\rm Ti}_2 \Big (pq,-\frac{q}{p} \Big ) - {\rm Ti}_2 \Big (\frac{p}{q},\frac{1}{pq} \Big ) - {\rm Ti}_2 \Big (\frac{p}{q},-\frac{1}{pq} \Big )  \Big ] \ ,
\eeqs
where $p=\sqrt{3}-\sqrt{2}$, $q=\sqrt{2}-1$, and Ti$_2(x,y)$ is the generalized inverse tangent integral \cite{invtan},
\beq
{\rm Ti}_2(x,y) = \int_0^x \frac{\tan^{-1} t}{t+y} \ dt \ .
\eeq

\subsubsection{Net 19}

Net 19 is shown in Fig. \ref{netfig2} (b), where a primitive unit cell contains nine vertices $\nu_{19}=9$, and the coordination number is $\kappa_{19}=4$. By the vertex labeling given in Fig. \ref{netfig2} (b), we have
\beq
M_{19}(\theta_1,\theta_2) = \left( \begin{array}{ccccccccc}
4 & -1 & -\beta & 0 & 0 & 0 & 0 & -\alpha & -1 \\
-1 & 4 & -1 & -1 & 0 & -\alpha & 0 & 0 & 0 \\ 
-\frac{1}{\beta} & -1 & 4 & -1 & 0 & -\alpha & 0 & 0 & 0 \\
0 & -1 & -1 & 4 & -1 & 0 & -1 & 0 & 0 \\
0 & 0 & 0 & -1 & 4 & -1 & 0 & -\frac{1}{\beta} & -\frac{1}{\beta} \\
0 & -\frac{1}{\alpha} & -\frac{1}{\alpha} & 0 & -1 & 4 & -1 & 0 & 0 \\
0 & 0 & 0 & -1 & 0 & -1 & 4 & -1 & -1 \\
-\frac{1}{\alpha} & 0 & 0 & 0 & -\beta & 0 & -1 & 4 & -1 \\
-1 & 0 & 0 & 0 & -\beta & 0 & -1 & -1 & 4
\end{array} \right ) \ . 
\eeq
The determinant can be calculated to be
\beqs
D_{19}(\theta_1,\theta_2) & = & 80 \{ 442 - 191 (\cos \theta_1 + \cos \theta_2) + 5 (\cos^2 \theta_1 + \cos^2 \theta_2) - 68 \cos \theta_1 \cos \theta_2 \cr\cr
& & - \cos \theta_1 \cos \theta_2 (\cos \theta_1 + \cos \theta_2) \} \ ,
\eeqs
such that
\beq
z_{19} = \frac{1}{9} \int_{-\pi}^{\pi} \frac{d\theta_1}{2\pi} \int_{-\pi}^{\pi} \frac{d\theta_2}{2\pi}
\ln \Big [ D_{19}(\theta_1,\theta_2) \Big ] = 1.144188002944693...  
\eeq

\subsubsection{Net 20}

Net 20 is shown in Fig. \ref{netfig2} (c), where a primitive unit cell contains ten vertices $\nu_{20}=10$, and the coordination number is $\kappa_{20}=4$. We have
\beq
M_{20}(\theta_1,\theta_2) = \left( \begin{array}{cccccccccc}
4 & -1 & 0 & -\beta & 0 & 0 & -\alpha\beta & 0 & -\alpha & 0 \\
-1 & 4 & -1 & 0 & 0 & -\beta & 0 & 0 & 0 & -1 \\
0 & -1 & 4 & -1 & -1 & 0 & 0 & 0 & -\alpha & 0 \\
-\frac{1}{\beta} & 0 & -1 & 4 & -1 & 0 & -\alpha & 0 & 0 & 0 \\
0 & 0 & -1 & -1 & 4 & -1 & 0 & -1 & 0 & 0 \\
0 & -\frac{1}{\beta} & 0 & 0 & -1 & 4 & -1 & 0 & 0 & -\frac{1}{\beta} \\
-\frac{1}{\alpha\beta} & 0 & 0 & -\frac{1}{\alpha} & 0 & -1 & 4 & -1 & 0 & 0 \\
0 & 0 & 0 & 0 & -1 & 0 & -1 & 4 & -1 & -1 \\
-\frac{1}{\alpha} & 0 & -\frac{1}{\alpha} & 0 & 0 & 0 & 0 & -1 & 4 & -1 \\
0 & -1 & 0 & 0 & 0 & -\beta & 0 & -1 & -1 & 4
\end{array} \right ) \ . 
\eeq
The determinant can be calculated to be
\beqs
D_{20}(\theta_1,\theta_2) & = & 16 \{ 7375 - 2995 (\cos \theta_1 + \cos \theta_2) + 34 (\cos^2 \theta_1 + \cos^2 \theta_2) - 1393 \cos \theta_1 \cos \theta_2 \cr\cr
& & - 30 \cos \theta_1 \cos \theta_2 (\cos \theta_1 + \cos \theta_2) \} \ ,
\eeqs
such that
\beq
z_{20} = \frac{1}{10} \int_{-\pi}^{\pi} \frac{d\theta_1}{2\pi} \int_{-\pi}^{\pi} \frac{d\theta_2}{2\pi}
\ln \Big [ D_{20}(\theta_1,\theta_2) \Big ] = 1.150677474300389...  
\eeq

\subsubsection{Net 21}

A primitive unit cell for net 21 containing twenty vertices $\nu_{21}=20$ is shown in Fig. \ref{netfig2} (d), and the coordination number is $\kappa_{21}=4$. Let us write $M_{21}(\theta_1,\theta_2) = M_{21}^\prime - M_{21}^{\prime\prime}(\theta_1,\theta_2)$, where $M_{21}^\prime$ is a diagonal matrix with all diagonal elements equal to four and 

\beq
M_{21}^{\prime\prime}(\theta_1,\theta_2)
= \left( \begin{array}{cccccccccccccccccccc}
0 & 1 & 0 & 1 & 0 & 0 & \beta & 0 & 0 & 0 & 0 & 0 & 0 & 0 & 1 & 0 & 0 & 0 & 0 & 0 \\
1 & 0 & 1 & 0 & 0 & 0 & \beta & 0 & 0 & 0 & \frac{\beta}{\alpha} & 0 & 0 & 0 & 0 & 0 & 0 & 0 & 0 & 0 \\
0 & 1 & 0 & 1 & 0 & 0 & 0 & 0 & \frac{\beta}{\alpha} & 0 & 0 & 0 & \frac{1}{\alpha} & 0 & 0 & 0 & 0 & 0 & 0 & 0 \\
1 & 0 & 1 & 0 & 1 & 0 & 0 & 0 & 0 & 0 & 0 & 0 & \frac{1}{\alpha} & 0 & 0 & 0 & 0 & 0 & 0 & 0 \\
0 & 0 & 0 & 1 & 0 & 1 & 0 & 0 & 0 & 0 & 0 & \frac{1}{\alpha} & 0 & 0 & 0 & 1 & 0 & 0 & 0 & 0 \\
0 & 0 & 0 & 0 & 1 & 0 & 1 & 0 & 0 & 0 & \frac{1}{\alpha} & 0 & 0 & 0 & 0 & 0 & 1 & 0 & 0 & 0 \\
\frac{1}{\beta} & \frac{1}{\beta} & 0 & 0 & 0 & 1 & 0 & 1 & 0 & 0 & 0 & 0 & 0 & 0 & 0 & 0 & 0 & 0 & 0 & 0 \\
0 & 0 & 0 & 0 & 0 & 0 & 1 & 0 & 1 & 0 & 0 & 0 & 0 & 0 & \frac{1}{\beta} & 0 & 0 & 1 & 0 & 0 \\
0 & 0 & \frac{\alpha}{\beta} & 0 & 0 & 0 & 0 & 1 & 0 & 1 & 0 & 0 & 0 & \frac{1}{\beta} & 0 & 0 & 0 & 0 & 0 & 0 \\
0 & 0 & 0 & 0 & 0 & 0 & 0 & 0 & 1 & 0 & 1 & 0 & 0 & 0 & 0 & 0 & 0 & 1 & 1 & 0 \\
0 & \frac{\alpha}{\beta} & 0 & 0 & 0 & \alpha & 0 & 0 & 0 & 1 & 0 & 1 & 0 & 0 & 0 & 0 & 0 & 0 & 0 & 0 \\
0 & 0 & 0 & 0 & \alpha & 0 & 0 & 0 & 0 & 0 & 1 & 0 & 1 & 0 & 0 & 0 & 0 & 0 & 1 & 0 \\
0 & 0 & \alpha & \alpha & 0 & 0 & 0 & 0 & 0 & 0 & 0 & 1 & 0 & 1 & 0 & 0 & 0 & 0 & 0 & 0 \\
0 & 0 & 0 & 0 & 0 & 0 & 0 & 0 & \beta & 0 & 0 & 0 & 1 & 0 & 1 & 0 & 0 & 0 & 0 & 1 \\
1 & 0 & 0 & 0 & 0 & 0 & 0 & \beta & 0 & 0 & 0 & 0 & 0 & 1 & 0 & 1 & 0 & 0 & 0 & 0 \\
0 & 0 & 0 & 0 & 1 & 0 & 0 & 0 & 0 & 0 & 0 & 0 & 0 & 0 & 1 & 0 & 1 & 0 & 0 & 1 \\
0 & 0 & 0 & 0 & 0 & 1 & 0 & 0 & 0 & 0 & 0 & 0 & 0 & 0 & 0 & 1 & 0 & 1 & 0 & 1 \\
0 & 0 & 0 & 0 & 0 & 0 & 0 & 1 & 0 & 1 & 0 & 0 & 0 & 0 & 0 & 0 & 1 & 0 & 1 & 0 \\
0 & 0 & 0 & 0 & 0 & 0 & 0 & 0 & 0 & 1 & 0 & 1 & 0 & 0 & 0 & 0 & 0 & 1 & 0 & 1 \\
0 & 0 & 0 & 0 & 0 & 0 & 0 & 0 & 0 & 0 & 0 & 0 & 0 & 1 & 0 & 1 & 1 & 0 & 1 & 0
\end{array} \right ) \ . 
\eeq
The determinant can be calculated to be
\beqs
D_{21}(\theta_1,\theta_2)
& = & 16 \{ 816275712 - 342967936 (\cos \theta_1 + \cos \theta_2) + 6766760 (\cos^2 \theta_1 + \cos^2 \theta_2) \cr\cr
& & - 139108816\cos \theta_1 \cos \theta_2 - 7200 (\cos^3 \theta_1 + \cos^3 \theta_2) \cr\cr
& & - 2389920 \cos \theta_1 \cos \theta_2 (\cos \theta_1 + \cos \theta_2) + (\cos^4 \theta_1 + \cos^4 \theta_2) \cr\cr
& & - 1160 \cos \theta_1 \cos \theta_2 (\cos^2 \theta_1 + \cos^2 \theta_2) + 32014\cos^2 \theta_1 \cos^2 \theta_2  \} \ ,
\eeqs
such that 
\beq
z_{21} = \frac{1}{20} \int_{-\pi}^{\pi} \frac{d\theta_1}{2\pi} \int_{-\pi}^{\pi} \frac{d\theta_2}{2\pi}
\ln \Big [ D_{21}(\theta_1,\theta_2) \Big ] = 1.155959257782222...
\eeq

\subsubsection{Net 22}

Net 22 contains only triangles and pentagons. Let us draw it as shown in Fig. \ref{netfig2} (e) with five vertices in a unit cell $\nu_{22}=5$. There are two $5^3$ and three $5^3.3$ vertices in each unit cell, so that the effective coordination number is $\kappa_{22}=18/5$. We have
\beq
M_{22}(\theta_1,\theta_2) = \left( \begin{array}{ccccc}
4 & -1 & -e^{i\theta_2} & -e^{i\theta_2} & -1 \\
-1 & 3 & -1 & -e^{-i\theta_1} & 0 \\
-e^{-i\theta_2} & -1 & 4 & -1 & -e^{-i(\theta_1+\theta_2)} \\
-e^{-i\theta_2} & -e^{i\theta_1} & -1 & 4 & -1 \\
-1 & 0 & -e^{i(\theta_1+\theta_2)} & -1 & 3
\end{array} \right ) \ . 
\eeq
The determinant can be calculated to be
\beqs
D_{22}(\theta_1,\theta_2) & = & 2 \{ 99 - 31 (\cos \theta_1 + \cos \theta_2 + \cos (\theta_1 + \theta_2)) \cr\cr
& & - 2 ( \cos \theta_1 \cos \theta_2 + (\cos \theta_1 + \cos \theta_2) \cos (\theta_1 + \theta_2) ) \} \ ,
\eeqs
such that
\beq
z_{22} = \frac{1}{5} \int_{-\pi}^{\pi} \frac{d\theta_1}{2\pi} \int_{-\pi}^{\pi} \frac{d\theta_2}{2\pi}
\ln \Big [ D_{22}(\theta_1,\theta_2) \Big ] = 1.024172110372259...  
\eeq

\subsubsection{Net 23}

A primitive unit cell for net 23 containing nine vertices $\nu_{23}=9$ is shown in Fig. \ref{netfig2} (f), and the coordination number is $\kappa_{23}=4$. We have
\beq
M_{23}(\theta_1,\theta_2) = \left( \begin{array}{ccccccccc}
4 & -1 & 0 & 0 & -\beta & -\beta & 0 & 0 & -1 \\
-1 & 4 & -1 & -1 & 0 & 0 & 0 & -\alpha\beta & 0 \\
0 & -1 & 4 & -1 & 0 & -\alpha\beta & 0 & 0 & -\alpha \\
0 & -1 & -1 & 4 & -1 & 0 & -1 & 0 & 0 \\
-\frac{1}{\beta} & 0 & 0 & -1 & 4 & -1 & 0 & -\alpha & 0 \\
-\frac{1}{\beta} & 0 & -\frac{1}{\alpha\beta} & 0 & -1 & 4 & -1 & 0 & 0 \\
0 & 0 & 0 & -1 & 0 & -1 & 4 & -1 & -1 \\
0 & -\frac{1}{\alpha\beta} & 0 & 0 & -\frac{1}{\alpha} & 0 & -1 & 4 & -1 \\
-1 & 0 & -\frac{1}{\alpha} & 0 & 0 & 0 & -1 & -1 & 4
\end{array} \right ) \ . 
\eeq
The determinant can be calculated to be
\beqs
D_{23}(\theta_1,\theta_2) & = & 8 \{ 4714 - 1473 (\cos \theta_1 + \cos \theta_2 + \cos (\theta_1 + \theta_2)) \cr\cr
& & - 103 ( \cos \theta_1 \cos \theta_2 + (\cos \theta_1 + \cos \theta_2) \cos (\theta_1 + \theta_2) ) \cr\cr
& & + 14 \cos \theta_1 \cos \theta_2 \cos (\theta_1 + \theta_2) \} \ ,
\eeqs
such that
\beq
z_{23} = \frac{1}{9} \int_{-\pi}^{\pi} \frac{d\theta_1}{2\pi} \int_{-\pi}^{\pi} \frac{d\theta_2}{2\pi}
\ln \Big [ D_{23}(\theta_1,\theta_2) \Big ] = 1.152329841150063...  
\eeq

\subsubsection{Net 24}

The coordination number for net 24 is $\kappa_{24}=4$.
The unit cell given in Fig. 26 of \cite{Okeeffe} contains ten vertices $\nu_{24}=10$. Using the vertex labeling given in the left-hand-side of Fig. \ref{netfig2} (g), we have
\beq
M_{24}(\theta_1,\theta_2) = \left( \begin{array}{cccccccccc}
4 & -1 & -\frac{1}{\beta} & 0 & 0 & -\frac{\alpha}{\beta} & -\alpha & 0 & 0 & 0 \\
-1 & 4 & -1 & 0 & 0 & 0 & -\alpha & 0 & -1 & 0 \\
-\beta & -1 & 4 & -1 & 0 & -\alpha & 0 & 0 & 0 & 0 \\
0 & 0 & -1 & 4 & -1 & 0 & 0 & 0 & -\beta & -1 \\
0 & 0 & 0 & -1 & 4 & -1 & 0 & -\beta & 0 & -1 \\
-\frac{\beta}{\alpha} & 0 & -\frac{1}{\alpha} & 0 & -1 & 4 & -1 & 0 & 0 & 0 \\
-\frac{1}{\alpha} & -\frac{1}{\alpha} & 0 & 0 & 0 & -1 & 4 & -1 & 0 & 0 \\
0 & 0 & 0 & 0 & -\frac{1}{\beta} & 0 & -1 & 4 & -1 & -1 \\
0 & -1 & 0 & -\frac{1}{\beta} & 0 & 0 & 0 & -1 & 4 & -1 \\
0 & 0 & 0 & -1 & -1 & 0 & 0 & -1 & -1 & 4
\end{array} \right ) \ , 
\eeq
with determinant 
\beqs
D_{24}(\theta_1,\theta_2)
& = & 4 \{ 30135 - 5929 \cos \theta_1 - 23101 \cos \theta_2 + 36 \cos^2 \theta_1 + 2456 \cos^2 \theta_2 \cr\cr
& & - 3341 \cos \theta_1 \cos \theta_2 - 4 \cos^3 \theta_2 - 248 \cos \theta_1 \cos^2  \theta_2 - 4 \cos \theta_1 \cos^3 \theta_2 \} \cr\cr
& = & 16 \Big ( 121 - 6 \cos^2 \frac{\theta_1}{2} - 52 \cos^2 \frac{\theta_2}{2} - 59 \cos \frac{\theta_1}{2} \cos \frac{\theta_2}{2} - 4 \cos \frac{\theta_1}{2} \cos^3 \frac{\theta_2}{2} \Big ) \cr\cr
& & \Big ( 121 - 6 \cos^2 \frac{\theta_1}{2} - 52 \cos^2 \frac{\theta_2}{2} + 59 \cos \frac{\theta_1}{2} \cos \frac{\theta_2}{2} + 4 \cos \frac{\theta_1}{2} \cos^3 \frac{\theta_2}{2} \Big ) \ .
\label{d24}
\eeqs
However, if one takes the primitive unit cell shown in the right-hand-side of Fig. \ref{netfig2} (g) containing five vertices, $\bar \nu_{24}=5$, we have
\beq
\bar M_{24}(\theta_1,\theta_2) = \left( \begin{array}{ccccc}
4 & -1 & -e^{i\theta_2} & -e^{i\theta_2} & -1 \\
-1 & 4 & -1 & -e^{-i\theta_1} & -1 \\
-e^{-i\theta_2} & -1 & 4 & -1 & -e^{-i(\theta_1+\theta_2)} \\
-e^{-i\theta_2} & -e^{i\theta_1} & -1 & 4 & -1 \\
-1 & -1 & -e^{i(\theta_1+\theta_2)} & -1 & 4
\end{array} \right ) \ ,
\eeq
with determinant 
\beqs
\bar D_{24}(\theta_1,\theta_2) & = & 2 \{ 184 - 61 \cos \theta_1 - 46 \cos \theta_2 - 61 \cos (\theta_1 + \theta_2) - 2 \cos \theta_1 \cos \theta_2 \cr\cr
& & - 12 \cos \theta_1 \cos (\theta_1 + \theta_2) - 2 \cos \theta_2 \cos (\theta_1 + \theta_2) \} \ .
\label{bard24}
\eeqs
Although the two determinants given in Eqs. (\ref{d24}) and (\ref{bard24}) are distinct, they give the same asymptotic growth constant for net 24,
\beqs
z_{24} & = & \frac{1}{10} \int_{-\pi}^{\pi} \frac{d\theta_1}{2\pi} \int_{-\pi}^{\pi} \frac{d\theta_2}{2\pi} \ln \Big [ D_{24}(\theta_1,\theta_2) \Big ] \cr\cr
& = & \frac{1}{5} \int_{-\pi}^{\pi} \frac{d\theta_1}{2\pi} \int_{-\pi}^{\pi} \frac{d\theta_2}{2\pi} \ln \Big [ \bar D_{24}(\theta_1,\theta_2) \Big ] 
= 1.148658687301195...  
\eeqs
This is an example of the integral identity one can obtain by choosing different unit cells in the calculation \cite{sti}.

\subsubsection{Net 25}

Let us draw net 25 as shown in Fig. \ref{netfig2} (h) with six vertices in a unit cell $\nu_{25}=6$, and the coordination number is $\kappa_{25}=4$. We have
\beq
M_{25}(\theta_1,\theta_2) = \left( \begin{array}{ccccccccc}
4 & -1 & 0 & -e^{i(\theta_2-\theta_1)} & -e^{-i\theta_1} & -1 \\
-1 & 4 & -1 & -e^{-i\theta_1} & 0 & -1 \\
0 & -1 & 4 & -1 & -1 & -e^{-i\theta_2} \\
-e^{i(\theta_1-\theta_2)} & -e^{i\theta_1} & -1 & 4 & -1 & 0 \\
-e^{i\theta_1} & 0 & -1 & -1 & 4 & -1 \\
-1 & -1 & -e^{i\theta_2} & 0 & -1 & 4
\end{array} \right ) \ .
\eeq
The determinant can be calculated to be
\beqs
D_{25}(\theta_1,\theta_2) & = & 4 \{ 295 - 104 \cos \theta_1 - 136 \cos \theta_2 + \cos^2 \theta_1 + 2 \cos^2 \theta_2 - 56 \cos \theta_1 \cos \theta_2 \cr\cr
& & - 2 \cos \theta_1 \cos^2 \theta_2 \} \ ,
\eeqs
such that
\beq
z_{25} = \frac{1}{6} \int_{-\pi}^{\pi} \frac{d\theta_1}{2\pi} \int_{-\pi}^{\pi} \frac{d\theta_2}{2\pi}
\ln \Big [ D_{25}(\theta_1,\theta_2) \Big ] = 1.150037457106072...  
\eeq

\subsection{Nets with heptagons, enneagons or octagons}

In this subsection ,we consider important nets which involve heptagons, enneagons or octagons.

\subsubsection{Net 26}

\begin{figure}[htbp]
\unitlength 0.8mm \hspace*{5mm}
\begin{picture}(170,55)
\put(15,0){\line(1,0){10}}
\put(25,0){\line(1,1){6}}
\put(15,0){\line(-1,1){6}}
\multiput(9,6)(-9,3){2}{\line(1,3){3}}
\multiput(9,6)(3,9){2}{\line(-3,1){9}}
\multiput(31,6)(-3,9){2}{\line(3,1){9}}
\multiput(31,6)(9,3){2}{\line(-1,3){3}}
\put(12,15){\line(2,1){8}}
\put(28,15){\line(-2,1){8}}
\put(20,19){\line(0,1){10}}
\multiput(15,0)(10,0){2}{\circle*{2}}
\multiput(9,6)(3,9){2}{\circle*{2}}
\multiput(0,9)(3,9){2}{\circle*{2}}
\multiput(31,6)(-3,9){2}{\circle*{2}}
\multiput(40,9)(-3,9){2}{\circle*{2}}
\multiput(20,19)(0,10){2}{\circle*{2}}
\put(22,31){\makebox(0,0){\footnotesize 1}}
\put(20,16){\makebox(0,0){\footnotesize 2}}
\put(11,18){\makebox(0,0){\footnotesize 3}}
\put(1,20){\makebox(0,0){\footnotesize 4}}
\put(-2,7){\makebox(0,0){\footnotesize 5}}
\put(7,4){\makebox(0,0){\footnotesize 6}}
\put(13,-2){\makebox(0,0){\footnotesize 7}}
\put(27,-2){\makebox(0,0){\footnotesize 8}}
\put(33,4){\makebox(0,0){\footnotesize 9}}
\put(43,7){\makebox(0,0){\footnotesize 10}}
\put(40,20){\makebox(0,0){\footnotesize 11}}
\put(30,18){\makebox(0,0){\footnotesize 12}}
\put(20,-7){\makebox(0,0){$(a)$}}

\multiput(60,15)(30,0){4}{\line(0,1){10}}
\multiput(60,15)(30,0){4}{\line(1,-2){5}}
\multiput(65,5)(30,0){3}{\line(2,-1){10}}
\multiput(90,15)(30,0){3}{\line(-1,-2){5}}
\multiput(85,5)(30,0){3}{\line(-2,-1){10}}
\multiput(85,5)(30,0){3}{\line(1,0){10}}
\multiput(75,40)(30,0){4}{\line(0,1){10}}
\multiput(75,40)(30,0){4}{\line(1,-2){5}}
\multiput(80,30)(30,0){3}{\line(2,-1){10}}
\multiput(75,40)(30,0){4}{\line(-1,-2){5}}
\multiput(70,30)(30,0){4}{\line(-2,-1){10}}
\multiput(70,30)(30,0){4}{\line(1,0){10}}
\multiput(95,55)(30,0){3}{\line(2,-1){10}}
\multiput(85,55)(30,0){3}{\line(-2,-1){10}}
\multiput(85,55)(30,0){3}{\line(1,0){10}}
\multiput(85,5)(10,0){2}{\circle*{2}}
\multiput(90,15)(0,10){2}{\circle*{2}}
\put(90,28){\makebox(0,0){\footnotesize 1}}
\put(88,17){\makebox(0,0){\footnotesize 2}}
\put(83,7){\makebox(0,0){\footnotesize 3}}
\put(97,7){\makebox(0,0){\footnotesize 4}}
\put(105,-7){\makebox(0,0){$(b)$}}
\end{picture}

\vspace*{15mm}

\begin{picture}(190,72)
\put(6,43){\line(1,0){10}}
\put(6,43){\line(-1,3){3}}
\put(16,43){\line(1,3){3}}
\put(3,52){\line(2,1){8}}
\put(19,52){\line(-2,1){8}}
\put(6,43){\line(-1,-1){6}}
\put(16,43){\line(1,-1){6}}
\put(0,37){\line(1,-3){3}}
\put(22,37){\line(-1,-3){3}}
\put(3,28){\line(2,-1){8}}
\put(19,28){\line(-2,-1){8}}
\put(19,28){\line(2,-1){8}}
\put(11,24){\line(-1,-3){3}}
\put(27,24){\line(1,-3){3}}
\put(8,15){\line(1,-1){6}}
\put(30,15){\line(-1,-1){6}}
\put(14,9){\line(1,0){10}}
\put(24,9){\line(1,-3){3}}
\put(11,56){\circle*{2}}
\multiput(3,52)(16,0){2}{\circle*{2}}
\multiput(6,43)(10,0){2}{\circle*{2}}
\multiput(0,37)(22,0){2}{\circle*{2}}
\multiput(3,28)(16,0){2}{\circle*{2}}
\multiput(11,24)(16,0){2}{\circle*{2}}
\multiput(8,15)(22,0){2}{\circle*{2}}
\multiput(14,9)(10,0){2}{\circle*{2}}
\put(27,0){\circle*{2}}
\put(25,-2){\makebox(0,0){\footnotesize 1}}
\put(22,7){\makebox(0,0){\footnotesize 2}}
\put(12,7){\makebox(0,0){\footnotesize 3}}
\put(6,13){\makebox(0,0){\footnotesize 4}}
\put(9,23){\makebox(0,0){\footnotesize 5}}
\put(1,30){\makebox(0,0){\footnotesize 6}}
\put(-2,39){\makebox(0,0){\footnotesize 7}}
\put(3,44){\makebox(0,0){\footnotesize 8}}
\put(1,54){\makebox(0,0){\footnotesize 9}}
\put(11,59){\makebox(0,0){\footnotesize 10}}
\put(22,54){\makebox(0,0){\footnotesize 11}}
\put(19,44){\makebox(0,0){\footnotesize 12}}
\put(25,39){\makebox(0,0){\footnotesize 13}}
\put(22,30){\makebox(0,0){\footnotesize 14}}
\put(30,23){\makebox(0,0){\footnotesize 15}}
\put(33,13){\makebox(0,0){\footnotesize 16}}
\put(15,-7){\makebox(0,0){$(c)$}}

\put(70,0){\line(0,1){10}}
\put(70,10){\line(-2,1){8}}
\put(70,10){\line(2,1){8}}
\put(62,14){\line(-1,3){3}}
\put(78,14){\line(1,3){3}}
\put(59,23){\line(1,1){6}}
\put(81,23){\line(-1,1){6}}
\put(65,29){\line(1,0){10}}
\put(65,29){\line(-1,3){3}}
\put(75,29){\line(1,3){3}}
\put(62,38){\line(2,1){8}}
\put(78,38){\line(-2,1){8}}
\put(70,42){\line(0,1){14}}
\put(70,56){\line(-2,1){8}}
\put(70,56){\line(2,1){8}}
\put(62,60){\line(1,3){3}}
\put(78,60){\line(-1,3){3}}
\put(65,69){\line(1,0){10}}
\put(78,38){\line(3,1){9}}
\put(78,60){\line(3,-1){9}}
\put(87,41){\line(1,2){4}}
\put(87,57){\line(1,-2){4}}
\put(62,38){\line(-3,1){9}}
\put(62,60){\line(-3,-1){9}}
\put(53,41){\line(-1,2){4}}
\put(53,57){\line(-1,-2){4}}
\put(53,41){\line(-1,-1){6}}
\put(47,35){\line(1,-3){3}}
\put(50,26){\line(3,-1){9}}
\multiput(70,0)(0,10){2}{\circle*{2}}
\multiput(62,14)(16,0){2}{\circle*{2}}
\multiput(59,23)(22,0){2}{\circle*{2}}
\multiput(65,29)(10,0){2}{\circle*{2}}
\multiput(62,38)(16,0){2}{\circle*{2}}
\multiput(70,42)(0,14){2}{\circle*{2}}
\multiput(62,60)(16,0){2}{\circle*{2}}
\multiput(65,69)(10,0){2}{\circle*{2}}
\multiput(53,41)(34,0){2}{\circle*{2}}
\multiput(49,49)(42,0){2}{\circle*{2}}
\multiput(53,57)(34,0){2}{\circle*{2}}
\multiput(47,35)(3,-9){2}{\circle*{2}}
\put(68,-2){\makebox(0,0){\footnotesize 1}}
\put(68,8){\makebox(0,0){\footnotesize 2}}
\put(60,12){\makebox(0,0){\footnotesize 3}}
\put(57,21){\makebox(0,0){\footnotesize 4}}
\put(48,24){\makebox(0,0){\footnotesize 5}}
\put(45,33){\makebox(0,0){\footnotesize 6}}
\put(51,41){\makebox(0,0){\footnotesize 7}}
\put(47,51){\makebox(0,0){\footnotesize 8}}
\put(51,59){\makebox(0,0){\footnotesize 9}}
\put(59,62){\makebox(0,0){\footnotesize 10}}
\put(62,71){\makebox(0,0){\footnotesize 11}}
\put(78,71){\makebox(0,0){\footnotesize 12}}
\put(81,62){\makebox(0,0){\footnotesize 13}}
\put(90,59){\makebox(0,0){\footnotesize 14}}
\put(94,51){\makebox(0,0){\footnotesize 15}}
\put(90,41){\makebox(0,0){\footnotesize 16}}
\put(81,36){\makebox(0,0){\footnotesize 17}}
\put(78,29){\makebox(0,0){\footnotesize 18}}
\put(84,21){\makebox(0,0){\footnotesize 19}}
\put(81,12){\makebox(0,0){\footnotesize 20}}
\put(62,29){\makebox(0,0){\footnotesize 21}}
\put(59,36){\makebox(0,0){\footnotesize 22}}
\put(67,44){\makebox(0,0){\footnotesize 23}}
\put(67,54){\makebox(0,0){\footnotesize 24}}
\put(70,-7){\makebox(0,0){$(d)$}}

\multiput(118,0)(26,0){3}{\line(1,0){10}}
\multiput(118,0)(26,0){3}{\line(-2,1){8}}
\multiput(128,0)(26,0){3}{\line(2,1){8}}
\multiput(110,4)(26,0){4}{\line(0,1){10}}
\multiput(118,18)(26,0){3}{\line(1,0){10}}
\multiput(118,18)(26,0){3}{\line(-2,-1){8}}
\multiput(128,18)(26,0){3}{\line(2,-1){8}}
\multiput(118,18)(26,0){3}{\line(-1,3){3}}
\multiput(128,18)(26,0){3}{\line(1,3){3}}
\multiput(115,27)(26,0){3}{\line(2,1){8}}
\multiput(131,27)(26,0){3}{\line(-2,1){8}}
\multiput(131,27)(26,0){2}{\line(1,0){10}}
\multiput(123,31)(26,0){3}{\line(0,1){10}}
\multiput(123,41)(26,0){3}{\line(-2,1){8}}
\multiput(123,41)(26,0){3}{\line(2,1){8}}
\multiput(115,45)(26,0){3}{\line(1,3){3}}
\multiput(131,45)(26,0){3}{\line(-1,3){3}}
\multiput(131,45)(26,0){2}{\line(1,0){10}}
\multiput(118,54)(26,0){3}{\line(1,0){10}}
\multiput(118,54)(26,0){3}{\line(-2,1){8}}
\multiput(128,54)(26,0){3}{\line(2,1){8}}
\multiput(110,58)(26,0){4}{\line(0,1){10}}
\multiput(110,68)(26,0){3}{\line(2,1){8}}
\multiput(136,68)(26,0){3}{\line(-2,1){8}}
\multiput(118,72)(26,0){3}{\line(1,0){10}}
\multiput(136,14)(0,44){2}{\circle*{2}}
\multiput(128,18)(16,0){2}{\circle*{2}}
\multiput(131,27)(10,0){2}{\circle*{2}}
\multiput(131,45)(10,0){2}{\circle*{2}}
\multiput(128,54)(16,0){2}{\circle*{2}}
\multiput(149,31)(0,10){2}{\circle*{2}}
\put(128,46){\makebox(0,0){\footnotesize 1}}
\put(126,56){\makebox(0,0){\footnotesize 2}}
\put(134,60){\makebox(0,0){\footnotesize 3}}
\put(146,56){\makebox(0,0){\footnotesize 4}}
\put(144,46){\makebox(0,0){\footnotesize 5}}
\put(147,39){\makebox(0,0){\footnotesize 6}}
\put(147,33){\makebox(0,0){\footnotesize 7}}
\put(144,26){\makebox(0,0){\footnotesize 8}}
\put(146,16){\makebox(0,0){\footnotesize 9}}
\put(133,12){\makebox(0,0){\footnotesize 10}}
\put(125,16){\makebox(0,0){\footnotesize 11}}
\put(127,26){\makebox(0,0){\footnotesize 12}}
\multiput(170,18)(10,0){2}{\circle*{2}}
\multiput(167,27)(16,0){2}{\circle*{2}}
\multiput(175,31)(0,10){2}{\circle*{2}}
\put(173,39){\makebox(0,0){\footnotesize 1}}
\put(173,33){\makebox(0,0){\footnotesize 2}}
\put(165,29){\makebox(0,0){\footnotesize 3}}
\put(167,19){\makebox(0,0){\footnotesize 4}}
\put(183,19){\makebox(0,0){\footnotesize 5}}
\put(185,29){\makebox(0,0){\footnotesize 6}}
\put(139,-7){\makebox(0,0){$(e)$}}
\end{picture}

\vspace*{5mm}
\caption{\footnotesize{(a) A unit cell of net 26. (b) Net 27. (c) A unit cell of net 28. (d) A unit cell of net 29. (e) Two unit cells of net 30. Vertices within a unit cell are labeled.}} 
\label{netfig3}
\end{figure}
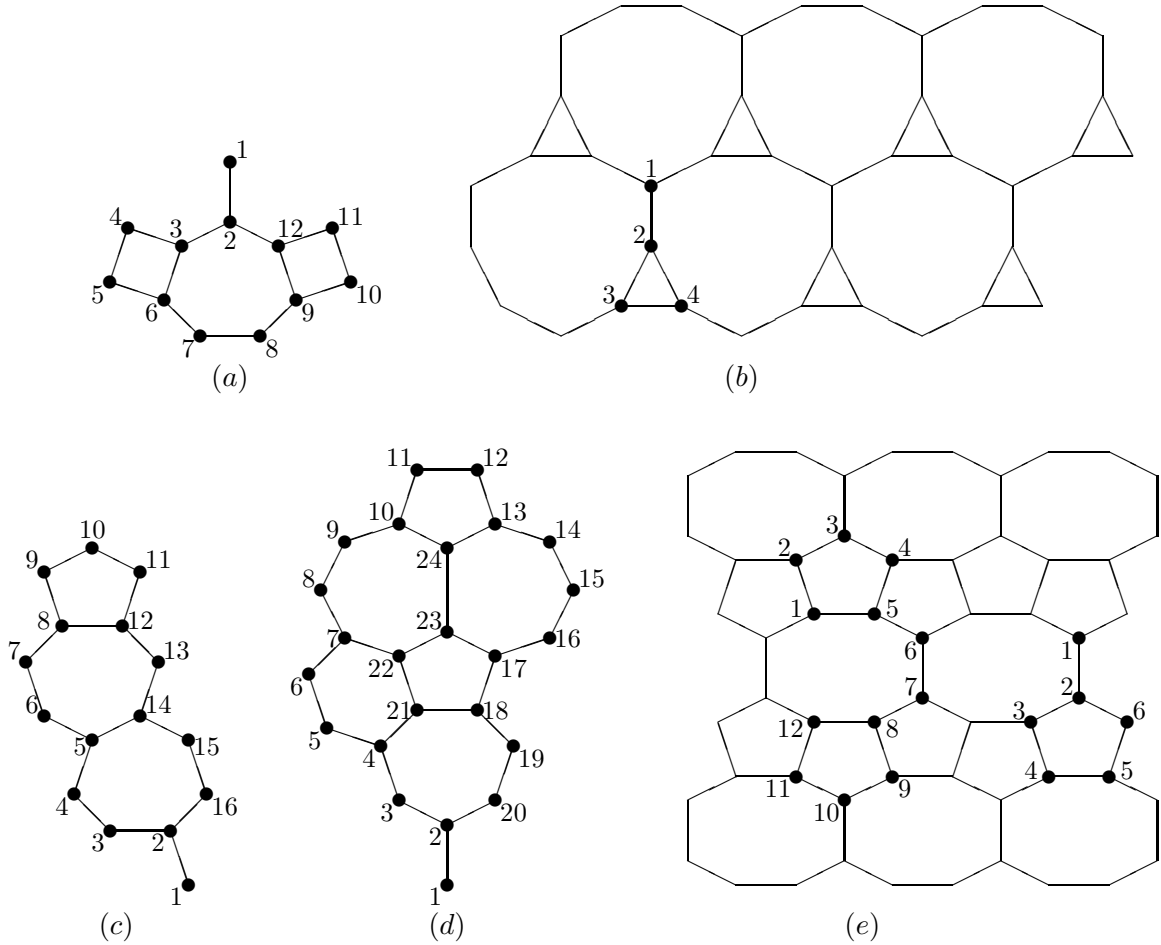

Net 26 contains only squares and heptagons. A primitive unit cell containing twelve vertices $\nu_{26}=12$ is shown in Fig. \ref{netfig3} (a), and the coordination number is $\kappa_{26}=3$. We have
\beq
M_{26}(\theta_1,\theta_2) = \left( \begin{array}{cccccccccccc}
3 & -1 & 0 & 0 & -\beta & 0 & 0 & 0 & 0 & -\frac{1}{\alpha} & 0 & 0 \\
-1 & 3 & -1 & 0 & 0 & 0 & 0 & 0 & 0 & 0 & 0 & -1 \\
0 & -1 & 3 & -1 & 0 & -1 & 0 & 0 & 0 & 0 & 0 & 0 \\
0 & 0 & -1 & 3 & -1 & 0 & 0 & -\frac{1}{\alpha} & 0 & 0 & 0 & 0 \\
-\frac{1}{\beta} & 0 & 0 & -1 & 3 & -1 & 0 & 0 & 0 & 0 & 0 & 0 \\
0 & 0 & -1 & 0 & -1 & 3 & -1 & 0 & 0 &  0 & 0 & 0 \\
0 & 0 & 0 & 0 & 0 & -1 & 3 & -1 & 0 & 0 & -\frac{1}{\beta} & 0 \\
0 & 0 & 0 & -\alpha & 0 & 0 & -1 & 3 & -1 & 0 & 0 & 0 \\
0 & 0 & 0 & 0 & 0 & 0 & 0 & -1 & 3 & -1 & 0 & -1 \\
-\alpha & 0 & 0 & 0 & 0 & 0 & 0 & 0 & -1 & 3 & -1 & 0 \\
0 & 0 & 0 & 0 & 0 & 0 & -\beta & 0 & 0 & -1 & 3 & -1 \\
0 & -1 & 0 & 0 & 0 & 0 & 0 & 0 & -1 & 0 & -1 & 3
\end{array} \right ) \ . 
\eeq
The determinant can be calculated to be
\beqs
D_{26}(\theta_1,\theta_2) & = & 4 \{ 4140 - 1752 (\cos \theta_1 + \cos \theta_2) + 37 (\cos^2 \theta_1 + \cos^2 \theta_2) - 686 \cos \theta_1 \cos \theta_2 \cr\cr
& & - 12 \cos \theta_1 \cos \theta_2 (\cos \theta_1 + \cos \theta_2) \} \ ,
\eeqs
such that
\beq
z_{26} = \frac{1}{12} \int_{-\pi}^{\pi} \frac{d\theta_1}{2\pi} \int_{-\pi}^{\pi} \frac{d\theta_2}{2\pi}
\ln \Big [ D_{26}(\theta_1,\theta_2) \Big ] = 0.7950402428757831...  
\eeq

\subsubsection{Net 27}

Net 27 contains only triangles and enneagons as shown in Fig. \ref{netfig3} (b), where each unit cell contains four vertices $\nu_{27}=4$, and the coordination number is $\kappa_{27}=3$. We have
\beq
M_{27}(\theta_1,\theta_2) = \left( \begin{array}{cccc}
3 & -1 & -e^{i\theta_2} & -e^{-i\theta_1} \\
-1 & 3 & -1 & -1 \\
-e^{-i\theta_2} & -1 & 3 & -1 \\
-e^{i\theta_1} & -1 & -1 & 3 
\end{array} \right ) \ . 
\eeq
The determinant can be calculated to be
\beq
D_{27}(\theta_1,\theta_2) = 8 \{ 3 - (\cos \theta_1 + \cos \theta_2) - \cos (\theta_1 + \theta_2)  \} \ ,
\eeq
which is four times of that for the triangular lattice \cite{wu77, sw, glasserwu}. It follows that
\beqs
z_{27} & = & \frac{1}{4} \int_{-\pi}^{\pi} \frac{d\theta_1}{2\pi} \int_{-\pi}^{\pi} \frac{d\theta_2}{2\pi}
\ln \Big [ D_{27}(\theta_1,\theta_2) \Big ] = \frac{1}{4} \Big ( z_{tri} + \ln 4 \Big ) \cr\cr
& = & \frac{3\sqrt{3}}{4\pi} \Big ( 1 - \frac{1}{5^2} + \frac{1}{7^2} - \frac{1}{11^2} + \frac{1}{13^2} - \cdots \Big ) + \frac{\ln 4}{4} \cr\cr
& = & 0.7504060243042857...  
\eeqs

\subsubsection{Net 28}

Net 28 is the B net of YCrB$_4$, which contains only pentagons and heptagons. A primitive unit cell containing sixteen vertices $\nu_{28}=16$ is shown in Fig. \ref{netfig3} (c), and the coordination number is $\kappa_{28}=3$. We have
\beq
M_{28}(\theta_1,\theta_2) = \left( \begin{array}{cccccccccccccccc}
3 & -1 & 0 & 0 & 0 & 0 & 0 & 0 & -\frac{\alpha}{\beta} & 0 & -\frac{1}{\beta} & 0 & 0 & 0 & 0 & 0 \\
-1 & 3 & -1 & 0 & 0 & 0 & 0 & 0 & 0 & 0 & 0 & 0 & 0 & 0 & 0 & -1 \\
0 & -1 & 3 & -1 & 0 & 0 & 0 & 0 & 0 & -\frac{1}{\beta} & 0 & 0 & 0 & 0 & 0 & 0 \\
0 & 0 & -1 & 3 & -1 & 0 & 0 & 0 & 0 & 0 & 0 & 0 & 0 & 0 & 0 & -\frac{1}{\alpha} \\
0 & 0 & 0 & -1 & 3 & -1 & 0 & 0 & 0 & 0 & 0 & 0 & 0 & -1 & 0 & 0 \\
0 & 0 & 0 & 0 & -1 & 3 & -1 & 0 & 0 & 0 & 0 & 0 & 0 & 0 & -\frac{1}{\alpha} & 0 \\
0 & 0 & 0 & 0 & 0 & -1 & 3 & -1 & 0 & 0 & 0 & 0 & -\frac{1}{\alpha} & 0 & 0 & 0 \\
0 & 0 & 0 & 0 & 0 & 0 & -1 & 3 & -1 & 0 & 0 & -1 & 0 & 0 & 0 & 0 \\
-\frac{\beta}{\alpha} & 0 & 0 & 0 & 0 & 0 & 0 & -1 & 3 & -1 & 0 & 0 & 0 & 0 & 0 & 0 \\
0 & 0 & -\beta & 0 & 0 & 0 & 0 & 0 & -1 & 3 & -1 & 0 & 0 & 0 & 0 & 0 \\
-\beta & 0 & 0 & 0 & 0 & 0 & 0 & 0 & 0 & -1 & 3 & -1 & 0 & 0 & 0 & 0 \\
0 & 0 & 0 & 0 & 0 & 0 & 0 & -1 & 0 & 0 & -1 & 3 & -1 & 0 & 0 & 0 \\
0 & 0 & 0 & 0 & 0 & 0 & -\alpha & 0 & 0 & 0 & 0 & -1 & 3 & -1 & 0 & 0 \\
0 & 0 & 0 & 0 & -1 & 0 & 0 & 0 & 0 & 0 & 0 & 0 & -1 & 3 & -1 & 0 \\
0 & 0 & 0 & 0 & 0 & -\alpha & 0 & 0 & 0 & 0 & 0 & 0 & 0 & -1 & 3 & -1 \\
0 & -1 & 0 & -\alpha & 0 & 0 & 0 & 0 & 0 & 0 & 0 & 0 & 0 & 0 & -1 & 3
\end{array} \right ) \ . 
\eeq
The determinant can be calculated to be
\beqs
D_{28}(\theta_1,\theta_2)
& = & 4 \{ 121795 - 115860 \cos \theta_1 - 11428 \cos \theta_2 + 19777 \cos^2 \theta_1 + 2 \cos^2 \theta_2 \cr\cr
& & - 11472 \cos \theta_1 \cos \theta_2 - 628 \cos^3 \theta_1 - 2132 \cos^2 \theta_1 \cos \theta_2 - 2 \cos \theta_1 \cos^2 \theta_2 \cr\cr
& & + 4 \cos^4 \theta_1 - 56 \cos^3 \theta_1 \cos \theta_2 \} \ ,
\eeqs
such that
\beq
z_{28} = \frac{1}{16} \int_{-\pi}^{\pi} \frac{d\theta_1}{2\pi} \int_{-\pi}^{\pi} \frac{d\theta_2}{2\pi}
\ln \Big [ D_{28}(\theta_1,\theta_2) \Big ] = 0.8025881723106822...  
\eeq

\subsubsection{Net 29}

Net 29 is the B net of Y$_2$LnB$_6$, which contains pentagons, hexagons and heptagons. A primitive unit cell containing twenty four vertices $\nu_{29}=24$ is shown in Fig. \ref{netfig3} (d), and the coordination number is $\kappa_{29}=3$. Let us write $M_{29}(\theta_1,\theta_2) = M_{29}^\prime - M_{29}^{\prime\prime}(\theta_1,\theta_2)$, where $M_{29}^\prime$ is a diagonal matrix with all diagonal elements equal to three and 
\beq
M_{29}^{\prime\prime}(\theta_1,\theta_2) = \left( \begin{array}{cccccccccccccccccccccccc}
0 & 1 & 0 & 0 & 0 & 0 & 0 & 0 & 0 & 0 & \frac{\alpha}{\beta} & 0 & 0 & 0 & \frac{1}{\beta} & 0 & 0 & 0 & 0 & 0 & 0 & 0 & 0 & 0 \\
1 & 0 & 1 & 0 & 0 & 0 & 0 & 0 & 0 & 0 & 0 & 0 & 0 & 0 & 0 & 0 & 0 & 0 & 0 & 1 & 0 & 0 & 0 & 0 \\
0 & 1 & 0 & 1 & 0 & 0 & 0 & 0 & 0 & 0 & 0 & 0 & 0 & \frac{1}{\beta} & 0 & 0 & 0 & 0 & 0 & 0 & 0 & 0 & 0 & 0 \\
0 & 0 & 1 & 0 & 1 & 0 & 0 & 0 & 0 & 0 & 0 & 0 & 0 & 0 & 0 & 0 & 0 & 0 & 0 & 0 & 1 & 0 & 0 & 0 \\
0 & 0 & 0 & 1 & 0 & 1 & 0 & 0 & 0 & 0 & 0 & \frac{1}{\beta} & 0 & 0 & 0 & 0 & 0 & 0 & 0 & 0 & 0 & 0 & 0 & 0 \\
0 & 0 & 0 & 0 & 1 & 0 & 1 & 0 & 0 & 0 & 0 & 0 & 0 & 0 & 0 & 0 & 0 & 0 & 0 & \frac{1}{\alpha} & 0 & 0 & 0 & 0 \\
0 & 0 & 0 & 0 & 0 & 1 & 0 & 1 & 0 & 0 & 0 & 0 & 0 & 0 & 0 & 0 & 0 & 0 & 0 & 0 & 0 & 1 & 0 & 0 \\
0 & 0 & 0 & 0 & 0 & 0 & 1 & 0 & 1 & 0 & 0 & 0 & 0 & 0 & 0 & 0 & 0 & 0 & \frac{1}{\alpha} & 0 & 0 & 0 & 0 & 0 \\
0 & 0 & 0 & 0 & 0 & 0 & 0 & 1 & 0 & 1 & 0 & 0 & 0 & 0 & 0 & \frac{1}{\alpha} & 0 & 0 & 0 & 0 & 0 & 0 & 0 & 0 \\
0 & 0 & 0 & 0 & 0 & 0 & 0 & 0 & 1 & 0 & 1 & 0 & 0 & 0 & 0 & 0 & 0 & 0 & 0 & 0 & 0 & 0 & 0 & 1 \\
\frac{\beta}{\alpha} & 0 & 0 & 0 & 0 & 0 & 0 & 0 & 0 & 1 & 0 & 1 & 0 & 0 & 0 & 0 & 0 & 0 & 0 & 0 & 0 & 0 & 0 & 0 \\
0 & 0 & 0 & 0 & \beta & 0 & 0 & 0 & 0 & 0 & 1 & 0 & 1 & 0 & 0 & 0 & 0 & 0 & 0 & 0 & 0 & 0 & 0 & 0 \\
0 & 0 & 0 & 0 & 0 & 0 & 0 & 0 & 0 & 0 & 0 & 1 & 0 & 1 & 0 & 0 & 0 & 0 & 0 & 0 & 0 & 0 & 0 & 1 \\
0 & 0 & \beta & 0 & 0 & 0 & 0 & 0 & 0 & 0 & 0 & 0 & 1 & 0 & 1 & 0 & 0 & 0 & 0 & 0 & 0 & 0 & 0 & 0 \\
\beta & 0 & 0 & 0 & 0 & 0 & 0 & 0 & 0 & 0 & 0 & 0 & 0 & 1 & 0 & 1 & 0 & 0 & 0 & 0 & 0 & 0 & 0 & 0 \\
0 & 0 & 0 & 0 & 0 & 0 & 0 & 0 & \alpha & 0 & 0 & 0 & 0 & 0 & 1 & 0 & 1 & 0 & 0 & 0 & 0 & 0 & 0 & 0 \\
0 & 0 & 0 & 0 & 0 & 0 & 0 & 0 & 0 & 0 & 0 & 0 & 0 & 0 & 0 & 1 & 0 & 1 & 0 & 0 & 0 & 0 & 1 & 0 \\
0 & 0 & 0 & 0 & 0 & 0 & 0 & 0 & 0 & 0 & 0 & 0 & 0 & 0 & 0 & 0 & 1 & 0 & 1 & 0 & 1 & 0 & 0 & 0 \\
0 & 0 & 0 & 0 & 0 & 0 & 0 & \alpha & 0 & 0 & 0 & 0 & 0 & 0 & 0 & 0 & 0 & 1 & 0 & 1 & 0 & 0 & 0 & 0 \\
0 & 1 & 0 & 0 & 0 & \alpha & 0 & 0 & 0 & 0 & 0 & 0 & 0 & 0 & 0 & 0 & 0 & 0 & 1 & 0 & 0 & 0 & 0 & 0 \\
0 & 0 & 0 & 1 & 0 & 0 & 0 & 0 & 0 & 0 & 0 & 0 & 0 & 0 & 0 & 0 & 0 & 1 & 0 & 0 & 0 & 1 & 0 & 0 \\
0 & 0 & 0 & 0 & 0 & 0 & 1 & 0 & 0 & 0 & 0 & 0 & 0 & 0 & 0 & 0 & 0 & 0 & 0 & 0 & 1 & 0 & 1 & 0 \\
0 & 0 & 0 & 0 & 0 & 0 & 0 & 0 & 0 & 0 & 0 & 0 & 0 & 0 & 0 & 0 & 1 & 0 & 0 & 0 & 0 & 1 & 0 & 1 \\
0 & 0 & 0 & 0 & 0 & 0 & 0 & 0 & 0 & 1 & 0 & 0 & 1 & 0 & 0 & 0 & 0 & 0 & 0 & 0 & 0 & 0 & 1 & 0
\end{array} \right ) \ . 
\eeq
The determinant can be calculated to be
\beqs
D_{29}(\theta_1,\theta_2)
& = & 8 \{ 36407767 - 21392474 \cos \theta_1 - 9913039 \cos \theta_2 + 1035241 \cos^2 \theta_1 \cr\cr
& & + 47642 \cos^2 \theta_2 - 5861590 \cos \theta_1 \cos \theta_2 - 4068 \cos^3 \theta_1 - 288591 \cos^2 \theta_1 \cos \theta_2 \cr\cr
& & - 31198 \cos \theta_1 \cos^2 \theta_2 - 2 \cos^3 \theta_2 + 2 \cos^4 \theta_1 - 840 \cos^3 \theta_1 \cos \theta_2 \cr\cr
& & + 1156 \cos^2 \theta_1 \cos^2 \theta_2 - 2 \cos \theta_1 \cos^3 \theta_2 - 4 \cos^3 \theta_1 \cos^2 \theta_2 \} \ , \cr & &
\eeqs
such that
\beq
z_{29} = \frac{1}{24} \int_{-\pi}^{\pi} \frac{d\theta_1}{2\pi} \int_{-\pi}^{\pi} \frac{d\theta_2}{2\pi}
\ln \Big [ D_{29}(\theta_1,\theta_2) \Big ] = 0.8043880179770491...  
\eeq

\subsubsection{Net 30}

Net 30 is a net with $5^2.8$ and $5.8^2$ vertices. Its coordination number is $\kappa_{30}=3$.
The unit cell given in Fig. 31 of \cite{Okeeffe} contains twelve vertices $\nu_{30}=12$. Using the vertex labeling given in the left-hand-side of Fig. \ref{netfig3} (e), we have
\beq
M_{30}(\theta_1,\theta_2) = \left( \begin{array}{cccccccccccc}
3 & -1 & 0 & 0 & -1 & -\frac{1}{\alpha} & 0 & 0 & 0 & 0 & 0 & 0 \\
-1 & 3 & -1 & -\frac{1}{\alpha} & 0 & 0 & 0 & 0 & 0 & 0 & 0 & 0 \\
0 & -1 & 3 & -1 & 0 & 0 & 0 & 0 & 0 & -\beta & 0 & 0 \\
0 & -\alpha & -1 & 3 & -1 & 0 & 0 & 0 & 0 & 0 & 0 & 0 \\
-1 & 0 & 0 & -1 & 3 & -1 & 0 & 0 & 0 & 0 & 0 & 0 \\
-\alpha & 0 & 0 & 0 & -1 & 3 & -1 & 0 & 0 & 0 & 0 & 0 \\
0 & 0 & 0 & 0 & 0 & -1 & 3 & -1 & 0 & 0 & 0 & -\alpha \\
0 & 0 & 0 & 0 & 0 & 0 & -1 & 3 & -1 & 0 & 0 & -1 \\
0 & 0 & 0 & 0 & 0 & 0 & 0 & -1 & 3 & -1 & -\alpha & 0 \\
0 & 0 & -\frac{1}{\beta} & 0 & 0 & 0 & 0 & 0 & -1 & 3 & -1 & 0 \\
0 & 0 & 0 & 0 & 0 & 0 & 0 & 0 & -\frac{1}{\alpha} & -1 & 3 & -1 \\
0 & 0 & 0 & 0 & 0 & 0 & -\frac{1}{\alpha} & -1 & 0 & 0 & -1 & 3
\end{array} \right ) \ , 
\eeq
with determinant 
\beqs
D_{30}(\theta_1,\theta_2)
& = & 4 \{ 5160 - 6791 \cos \theta_1 - 169 \cos \theta_2 + 2261 \cos^2 \theta_1 - 221 \cos \theta_1 \cos \theta_2 \cr\cr
& & - 184 \cos^3 \theta_1 - 56 \cos^2 \theta_1 \cos  \theta_2 + 4 \cos^4 \theta_1 - 4 \cos^3 \theta_1 \cos \theta_2 \} \cr\cr
& = & 16 \Big ( 60 - 49 \cos^2 \frac{\theta_1}{2} + 11 \cos \frac{\theta_1}{2} \cos \frac{\theta_2}{2} + 4 \cos^4 \frac{\theta_1}{2}  + 4 \cos^3 \frac{\theta_1}{2} \cos \frac{\theta_2}{2} \Big ) \cr\cr
& & \Big ( 60 - 49 \cos^2 \frac{\theta_1}{2} - 11 \cos \frac{\theta_1}{2} \cos \frac{\theta_2}{2} + 4 \cos^4 \frac{\theta_1}{2} - 4 \cos^3 \frac{\theta_1}{2} \cos \frac{\theta_2}{2} \Big ) \ .
\label{d30}
\eeqs
However, if one takes the primitive unit cell shown in the right-hand-side of Fig. \ref{netfig3} (e), where each unit cell contains six vertices, $\bar \nu_{30}=6$, we have
\beq
\bar M_{30}(\theta_1,\theta_2) = \left( \begin{array}{cccccc}
3 & -1 & 0 & -e^{i\theta_2} & -e^{i(\theta_2-\theta_1)} & 0 \\
-1 & 3 & -1 & 0 & 0 & -1 \\
0 & -1 & 3 & -1 & 0 & -e^{-i\theta_1} \\
-e^{-i\theta_2} & 0 & -1 & 3 & -1 & 0 \\
-e^{i(\theta_1-\theta_2)} & 0 & 0 & -1 & 3 & -1 \\
0 & -1 & -e^{i\theta_1} & 0 & -1 & 3
\end{array} \right ) \ ,
\eeq
with determinant 
\beqs
\bar D_{30}(\theta_1,\theta_2) & = & 2 \{ 73 - 45 \cos \theta_1 - 13 \cos \theta_2 - 13 \cos (\theta_1 - \theta_2) + 2 \cos^2 \theta_1 - 2 \cos \theta_1 \cos \theta_2 \cr\cr
& & - 2 \cos \theta_1 \cos (\theta_1 - \theta_2) \} \ .
\label{bard30}
\eeqs
Although the two determinants given in Eqs. (\ref{d30}) and (\ref{bard30}) are distinct, they give the same asymptotic growth constant for net 30,
\beqs
z_{30} & = & \frac{1}{12} \int_{-\pi}^{\pi} \frac{d\theta_1}{2\pi} \int_{-\pi}^{\pi} \frac{d\theta_2}{2\pi} \ln \Big [ D_{30}(\theta_1,\theta_2) \Big ] \cr\cr
& = & \frac{1}{6} \int_{-\pi}^{\pi} \frac{d\theta_1}{2\pi} \int_{-\pi}^{\pi} \frac{d\theta_2}{2\pi} \ln \Big [ \bar D_{30}(\theta_1,\theta_2) \Big ] 
= 0.7985013545791521...  
\eeqs
This is another example of the integral identity one can obtain by choosing different unit cells in the calculation \cite{sti}.

\section{Discussion}

It is of interest to see how close the exact results presented above are to the upper bound given in Eq. (\ref{mcybound}).  For this purpose, we define the ratio
\beq
r_{\Lambda_k} = \frac{z_{\Lambda_k}}{\ln b_k} 
\label{rupper}
\eeq
for a $k$-regular lattice $\Lambda_k$, where $b_k$ is given by Eq. (\ref{ck}). For a lattice $\Lambda$ which is not $k$-regular, we replace $k$ by $\kappa$ in Eq. (\ref{ck}) and consider the ratio
\beq
r_\Lambda = \frac{z_\Lambda}{\ln b_\kappa} \ .
\label{ruppern}
\eeq

The values of $z_\Lambda$ and $r_\Lambda$ for various lattices $\Lambda$ are summarized in Table \ref{ztable}. Our results agree with the observations made in \cite{sw} that $z_\Lambda$ is relatively large for large value of $k$ (or $\kappa$). For the two-dimensional lattices with $k=4$ studied here and in \cite{sw, sti}, their values of $z_\Lambda$ are all smaller than that of the square lattice, $z_{sq}=4C/\pi=1.166243616123275...$ \cite{wu77}, which indicates that the square lattice may be the most densely connected two-dimensional lattice with $k=4$ \cite{sw}. Similarly, the values of $z_\Lambda$ for the lattices with $k=3$ studied here and in \cite{sw, sti} are smaller than that of the honeycomb lattice, $z_{hc}=0.8076648680486262...$ \cite{wu77}, which indicates that the honeycomb lattice may be the most densely connected two-dimensional lattice with $k=3$. The triangular lattice, the dual of the honeycomb lattice, would be the most densely connected two-dimensional lattice with $k=6$.

The ratios $r_\Lambda$ are close to each other no matter the lattice $\Lambda$ is $k$-regular or not. We thereby conjecture the upper bound 
\beq
z_{\Lambda} \leq \ln(b_\kappa)
\eeq
for a lattice $\Lambda$ with effective coordination number $\kappa \ge 3$, which generalizes the bound given in Eq. (\ref{mcybound}) for $k$-regular lattices.

\begin{table}
\caption{\label{ztable} Number of vertices in a primitive unit cell $\nu_\Lambda$, effective coordination number $\kappa_\Lambda$, and numerical values of $z_\Lambda$ and $r_\Lambda$. The last digits given in the text are rounded off.}
\begin{center}
\begin{tabular}{|c|c|c|l|l|}
\hline
$\Lambda$& $\nu_\Lambda$ & $\kappa_\Lambda$ & $z_\Lambda$       & $r_\Lambda$ \\
\hline
net 12   &            12 &                5 & 1.40973790375693  & 0.950533252598831 \\ 
net 13   &             8 &                5 & 1.40913328642468  & 0.950125581869492 \\ 
net 14   &             3 &                4 & 1.12777836380554  & 0.927147894482279 \\ 
net 15   &            11 &  $\frac{42}{11}$ & 1.07327025442306  & 0.926486463318153 \\ 
net 16a  &            12 &        $\frac92$ & 1.28028724864248  & 0.941934677506853 \\ 
net 16b  &            12 &        $\frac92$ & 1.27761792670833  & 0.939970800339543 \\ 
net 17   &            28 &   $\frac{32}{7}$ & 1.29917775354410  & 0.942842965374234 \\ 
net 18   &             6 &   $\frac{10}{3}$ & 0.940570430496223 & 0.957234196587022 \\ 
net 19   &             9 &                4 & 1.14418800294469  & 0.940638277757358 \\ 
net 20   &            10 &                4 & 1.15067747430039  & 0.945973279648539 \\ 
net 21   &            20 &                4 & 1.15595925778222  & 0.950315439944796 \\ 
net 22   &             5 &   $\frac{18}{5}$ & 1.02417211037226  & 0.945174625646314 \\ 
net 23   &             9 &                4 & 1.15232984115006  & 0.947331692342695 \\ 
net 24   &             5 &                4 & 1.14865868730120  & 0.944313632526569 \\ 
net 25   &             6 &                4 & 1.15003745710607  & 0.945447120774431 \\ 
net 26   &            12 &                3 & 0.795040242875783 & 0.949882240790510 \\ 
net 27   &             4 &                3 & 0.750406024304286 & 0.896555064043746 \\ 
net 28   &            16 &                3 & 0.802588172310682 & 0.958900204584415 \\
net 29   &            24 &                3 & 0.804388017977049 & 0.961050588102690 \\
net 30   &             6 &                3 & 0.798501354579152 & 0.954017438436016 \\
\hline
\end{tabular}
\end{center}
\end{table}

\section{Acknowledgments}
The author thanks Prof. J. Kozak for helpful discussion. This research was partially supported by the Taiwan NSC grant NSC-97-2112-M-006-007-MY3 and NSC-97-2119-M-002-001.

\end{document}